\documentclass[sigplan,nonacm]{acmart}

\makeatletter
\def\@ACM@checkaffil{
    \if@ACM@instpresent\else
    \ClassWarningNoLine{\@classname}{No institution present for an affiliation}%
    \fi
    \if@ACM@citypresent\else
    \ClassWarningNoLine{\@classname}{No city present for an affiliation}%
    \fi
    \if@ACM@countrypresent\else
        \ClassWarningNoLine{\@classname}{No country present for an affiliation}%
    \fi
}
\makeatother

\usepackage{amsmath,amsfonts}
\usepackage{graphicx}
\usepackage{textcomp}
\usepackage{xcolor}
\usepackage{algorithm,algpseudocode}
\usepackage{booktabs}
\usepackage{xspace}
\usepackage{soul}
\usepackage{enumitem}

\def\invcircledast#1{%
  \mathbin{\vphantom{\circledast}\text{%
    \ooalign{\smash{\blackcircle}\cr
             \hidewidth\smash{\textcolor{white}{\bf \footnotesize $#1$}}\hidewidth\cr
            }%
  }}%
}

\newcommand{\blackcircle}{\raisebox{-.6ex}{\scalebox{2.30}{$\bullet$}}}
\newcommand{\Design}{$\mathsf{HDReason}$\xspace}

\begin{document}
\title{HDReason: Algorithm-Hardware Codesign for Hyperdimensional Knowledge Graph Reasoning}

\author{Hanning Chen$^1$, Yang Ni$^1$, Ali Zakeri$^1$, Zhuowen Zou$^1$, Sanggeon Yun$^1$, Fei Wen$^2$, Behnam Khaleghi$^3$, Narayan Srinivasa$^4$, Hugo Latapie$^5$, and Mohsen Imani$^1$} 
\affiliation{$^1$Department of Computer Science, University of California, Irvine, CA, USA \\
             $^2$Department of Electrical and Computer Engineering, Texas A\&M University, TX, USA \\
             $^3$Qualcomm, CA, USA \\
             $^4$Intel Labs, CA, USA \\
             $^5$Cisco Systems, CA, USA \\          
$^*$ Corresponding author: m.imani@uci.edu}  

\begin{abstract}
In recent times, a plethora of hardware accelerators have been put forth for graph learning applications such as vertex classification and graph classification. However, previous works have paid little attention to Knowledge Graph Completion (KGC), a task that is well-known for its significantly higher algorithm complexity.
The state-of-the-art KGC solutions based on graph convolution neural network (GCN) involve extensive vertex/relation embedding updates and complicated score functions, which are inherently cumbersome for acceleration. As a result, existing accelerator designs are no longer optimal, and a novel algorithm-hardware co-design for KG reasoning is needed.

Recently, brain-inspired HyperDimensional Computing (HDC) has been introduced as a promising solution for lightweight machine learning, particularly for graph learning applications. In this paper, we leverage HDC for an intrinsically more efficient and acceleration-friendly KGC algorithm. We also co-design an acceleration framework named \Design targeting FPGA platforms. On the algorithm level, \Design achieves a balance between high reasoning accuracy, strong model interpretability, and less computation complexity. In terms of architecture, \Design offers reconfigurability, high training throughput, and low energy consumption. 
When compared with NVIDIA RTX 4090 GPU, the proposed accelerator achieves an average \textbf{10.6$\times$} speedup and \textbf{65$\times$} energy efficiency improvement. 
When conducting cross-models and cross-platforms comparison, \Design yields an average \textbf{4.2$\times$} higher performance and \textbf{3.4$\times$} better energy efficiency with similar accuracy versus the state-of-the-art FPGA-based GCN training platform.
 
\end{abstract}

\maketitle
\pagestyle{plain}


\section{Introduction}

Knowledge Graph (KG), usually on massive scales~\cite{toutanova2015observed,dettmers2018convolutional,miller1995wordnet,mahdisoltani2014yago3}, helping computers understand the relationship between each item of data. 
Each pair of connected vertices and the edge in between can be organized into a single triple that indicates a certain relation between two entities (also known as a fact). 
Knowledge Graph Completion (KGC), as a fundamental KG reasoning technique, infers new knowledge from incomplete facts by leveraging the semantics embedded in the graph structure. 
Therefore, it reasons upon vast arrays of information in a structured and meaningful way, crucial to various domains including question-answering, recommendation systems, and drug discovery~\cite{huang2019knowledge,zhang2018variational,wang2019knowledge,guo2020survey,zeng2022toward}.
Despite being widely applied for graph reasoning, KGC poses significant challenges, both algorithmically and in terms of hardware acceleration.

From an algorithmic perspective, existing algorithms, nam-ely the embedding, Graph Neural Network (GNN), and Reinforcement Learning (RL) based methods, all have their own drawbacks. 
Embedding-based techniques often yield low-quality results due to the limited representational capacity of their model~\cite{bordes2013translating,yang2015embedding}. 
GNN-based methods are specially designed to extract features from graph structure~\cite{vashishth2019composition,schlichtkrull2018modeling,shang2019end}, powerful in discovering more underlying semantics for higher quality reasoning. However, due to the extensive use of multi-hop and nonlinear vertex aggregation, they are computationally heavy, less transparent, and incur costly inference and training overheads when deployed on conventional hardware. 
RL-based approaches~\cite{xiong2017deeppath, stoica2020contextual, hildebrandt2020reasoning, wu2021findings, wang2020adrl} suffer from long latency and stability issues due to the exploration-exploitation dilemma of RL, making them less desirable for real-world applications~\cite{hildebrandt2020reasoning}. They are also limited to single-direction reasoning only (as defined in Section~\ref{sec:KG_basic})~\cite{wu2021findings, wang2020adrl}. 

The hardware accelerator challenges are equally daunting. As the amount of data grows, the size of KGs grows substantially and thereby necessitates hardware acceleration to make any existing KGC algorithm feasible.
GPU platforms~\cite{wang2021gnnadvisor,yang2022gnnlab,yan2020characterizing} have always been a go-to choice for implementing and verifying new algorithms. However, deploying GNN-based approaches on GPUs (or TPUs) introduces a significant memory bottleneck. Challenges in GPU acceleration also arise with RL-based methods due to the training and inference being highly intertwined, which leads to large off-chip data traffic and limited parallelism at the algorithmic level~\cite{cho2019fa3c}.
On the other hand, FPGAs can take advantage of customization when co-designing the algorithm and hardware, thereby exploiting more parallelism in the algorithm, improving the overall energy efficiency, and reducing the substantial memory traffic~\cite{geng2020awb,sarkar2023flowgnn}.

Unfortunately, current FPGA-based accelerators face multiple challenges when targeting KGC applications~\cite {sarkar2023flowgnn,geng2020awb,zeng2020graphact,lin2022hp}. 
\underline{First}, the vertex and edge embedding training play a pivotal role during the KGC process due to natural semantic attributes in KG. Existing FPGA accelerators for GNN-based methods, however, only focus on GNN propagation weight training. 
\underline{Second}, most existing designs, including both FPGA and non-FPGA based~\cite{kiningham2022grip,geng2020awb,yan2020hygcn}, rely on matrix-matrix multiplication accelerations which make it hard to support models that involve edge embedding. For example, the State-of-the-art design that supports edge-embedding (i.e., FlowGNN~\cite{sarkar2023flowgnn}) only supports inference instead of training. 
\underline{Third}, GNN and RL-based algorithms are not robust to quantization where model quality significantly deteriorates, limiting their usability in resource-constrained platforms like FPGA.
To build successful KGC acceleration platforms based on FPGA, all three challenges need to be resolved. 
\textbf{Therefore we need to redesign both algorithms and hardware acceleration to build an efficient and effective KGC framework.}

In light of these challenges, Hyperdimensional Computing (HDC) emerges as a promising solution. 
Motivated by human brains that process, represent, and memorize information by high-dimensional neural signals, HDC is centered around the distributed high-dimensional vector representation, i.e., hypervectors~\cite{ge2020classification,karunaratne2020memory}. These hypervectors are known to be holographic, meaning that information is evenly encoded in all vector elements, providing robustness to model quantization and representation redundancy.
This holographicness also allows HDC to represent and manipulate both symbolic and numerical data in a transparent and interpretable way, with efficient and natural memorization and reasoning in its model~\cite{kleyko2023survey,zou2022biohd,imani2017voicehd, ni2022neurally}.
Unlike traditional Deep Neural Network (DNN) approaches, HDC is inherently hardware-friendly, especially for customizable hardware platforms like FPGAs~\cite{Hanning_ICCAD2022,F5-HD,imani2021revisiting,zou2021scalable,imani2019sparsehd,ni2023brain}. The HDC model easily attains computation parallelism along hypervector directions due to its holographic nature. Moreover, the HDC model can achieve high learning accuracy at low bit precision, facilitating its mapping into FPGA computing logic. 

These advantages help recent works in HDC~\cite{poduval2022graphd,nunes2022graphhd,kang2022relhd} to achieve transparent and efficient graph reconstruction, node classification, and graph matching; manifesting its potential for other learning and reasoning tasks over graphs.
While HDC has been widely applied, it is not without its limitations, especially when deployed on traditional GPU/CPU platforms. The limited computation parallelism of CPUs restricts the HDC model's learning throughput. The GPU's limited on-chip memory and fixed datapath increase hypervector storage costs, thereby reducing memory efficiency. Additionally, for large-scale graph reasoning tasks, the GPU's power efficiency is low. These limitations typically involve issues of scalability and efficiency in handling large-scale graph reasoning tasks. Furthermore, previous attempts to accelerate HDC on FPGAs have not explicitly considered the stringent demands of large-scale graph reasoning on computing and memory resources, leading to suboptimal solutions~\cite{imani2021revisiting,salamat2020accelerating}. Specifically, previous HDC FPGA accelerator did not consider processing large-scale graph dataset with high sparsity and computation imbalance. Considering the large gap between the size of knowledge graph dataset and FPGA on-chip storage, we need to design customized datapath and scheduling algorithm to achieve high reasoning throughput.

To address these gaps, our work presents a novel KGC algorithm based on HDC, leveraging its intriguing properties for efficient and effective reasoning.
We explore an algorithm-hardware codesign approach, combining the strengths of FPGA and HDC, to create an energy and resource-efficient graph reasoning framework named \Design, which fully supports and accelerates the end-to-end KGC. The main contributions in this paper are listed as follows:
\begin{itemize}[leftmargin=*]
    \item To the best of our knowledge, \Design is the first FPGA-based framework for KGC acceleration. \Design is also the first neurosymbolic HDC model targeting KGC applications, by leveraging hypervector operations for vertex and relation embedding vector encoding and transparent vertex neighbor information memorization.
    
    \item On the algorithm part, \Design uses brain-inspired HDC to reduce computation complexity while maintaining high reasoning accuracy and strong model interpretability. Thanks to the HDC model's transparency, \Design model is capable of discovering underlying semantics without bulky computation and training. HDC model's high parallelism and low computation complexity also lift previous limitations on accelerator architecture design.
    
    \item On the hardware side, targeting \Design model, we propose several hardware optimizations including reusing encoded hypervectors, balancing vertex-to-vertex computation, and computing training backward gradients in the forward path (i.e., forward/backward co-optimizations). The proposed tuning techniques help \Design achieve significant performance improvement compared to existing hardware platforms.

    \item Through hardware and software codesign, the HDReason model supports low-bit precision while maintaining high reasoning accuracy. Its robustness is evident in maintaining accuracy even with limited hypervector dimensions during the reasoning process.
\end{itemize}

We evaluate \Design performance on several large-scale KG datasets. 
In terms of reasoning efficiency, \Design provides on average 4.2$\times$(and 8.3$\times$) speedup and 3.4$\times$ (and 9.1$\times$) energy efficiency improvement with similar accuracy as compared to the state-of-the-art GCN training platform HP-GNN~\cite{lin2022hp} (and GraphACT~\cite{zeng2020graphact}). 
When compared with \textbf{RTX 4090 GPU}, our accelerator achieves on average \textbf{10.6$\times$} speedup and \textbf{65$\times$} energy efficiency improvement.

\section{Background}

\subsection{HDC Basics}
\begin{figure}[t!]
  \centering
  \includegraphics[width=1\linewidth]{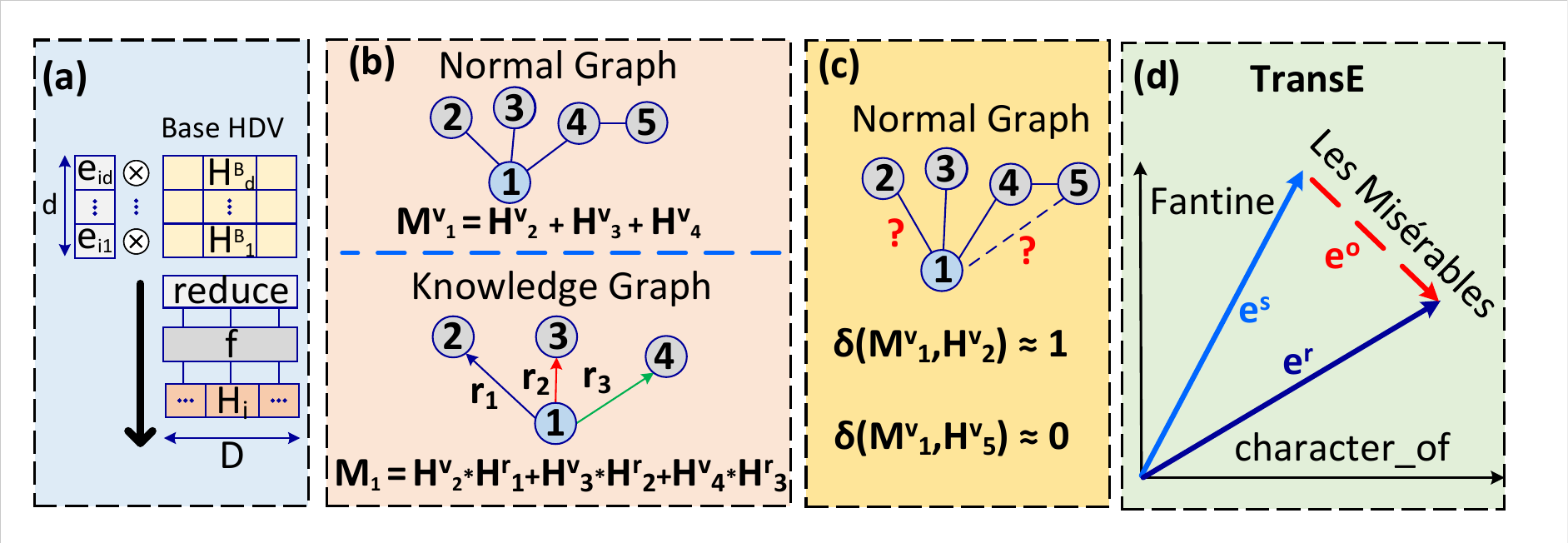}
  \caption{(a) HDC encoding example. (b) HDC memorization in graph learning. (c) Vertex neighbor reconstruction. (d) Score function example, TransE.}
  \vspace{-3mm}
  \label{Fig:HDReason_base}
\end{figure}

\noindent \textbf{HDC Encoding} As a foundation of HDC, it maps data from the original space into the high-dimensional space (i.e., hyperspace)~\cite{imani2020dual,montagna2018pulp}. Among a variety of HDC encoding methods~\cite{ge2020classification}, we choose the state-of-the-art kernel-based HDC encoding~\cite{imani2020dual}. It is inspired by kernel tricks, where the dot products between encoded hypervectors approximate values of a certain kernel. Suppose the original dimension of the data is $d$ and the hyperspace dimension is $D$, as is shown in Figure~\ref{Fig:HDReason_base}(a). The mathematical equation of the encoding process is: $\vec{H} = tanh(\vec{e}\text{ }\mathbf{H^B})$, where $\vec{e}$ is the original data with dimension $1 \times d$ and $\textbf{H}^\textbf{B}$ is a matrix of base hypervectors (Base HDV) with dimension $d \times D$. Each element of the Base HDV is generated randomly using the standard Gaussian distribution $\mathcal{N}(0,1)$ and stays constant. 

\noindent \textbf{Hypervector Operations} 
HDC has a set of well-defined hypervector operations, with the most important two being: (1) \textit{Bundling} (or element-wise addition "$+$"), which represents and memorizes a set of hypervectors. (2) \textit{Binding} (or element-wise multiplication "$\circ$"), which associates hypervectors for different concepts or symbols.

\noindent \textbf{Graph Structure Memorization} Bundling and binding operations give HDC the ability to represent rich combinations of symbols with hypervectors of the same size as each constituent, and more importantly, to efficiently memorize and later retrieve those symbols. Specifically in graph-related tasks, prior works utilize HDC to memorize each vertex's neighbor information and perform graph learning~\cite{poduval2022graphd,nunes2022graphhd,kang2022relhd}. Compared to GCN, these solutions serve as efficient, robust, and scalable alternatives. Figure~\ref{Fig:HDReason_base}(b) illustrates an example of memorizing neighbor information into a single hypervector. Suppose we have vertex $i$ and its neighbor vertex is represented as $N(i)$. Then we have the mathematical equation of vertex neighbor memorization as:
\begin{equation} \label{eq:memorization_base}
     \vec{M^v_{i}} = \Sigma_{j\in N(i)} \Vec{H_j^v}
\end{equation}
Here $\vec{M_{i}}$ is the memory hypervector of vertex $i$.

\noindent \textbf{Vertex Neighbor Reconstruction} After generating memory hypervectors for all vertices, we can perform several different learning tasks such as node classification and graph matching~\cite{nunes2022graphhd,poduval2022graphd}, similar to DNN based model like GNN. Furthermore, a property unique to HDC is that the neighbor information of each vertex can be reconstructed~\cite{poduval2022graphd}, specifically, we can determine whether two vertices are connected. As is shown in Figure~\ref{Fig:HDReason_base} (c), to test whether two vertices are connected, we have:
\begin{equation}
    R_{ij} = \delta (\vec{M^v_i}, \vec{H^v_j})
\end{equation}
Here $R_{ij}$ is the possibility that vertex $i$ and vertex $j$ are connected. $\delta$ is the vector distance measuring function such as Hamming distance function or cosine similarity function.

\begin{figure*}[t!]
  \centering
  \includegraphics[width=0.9\linewidth]{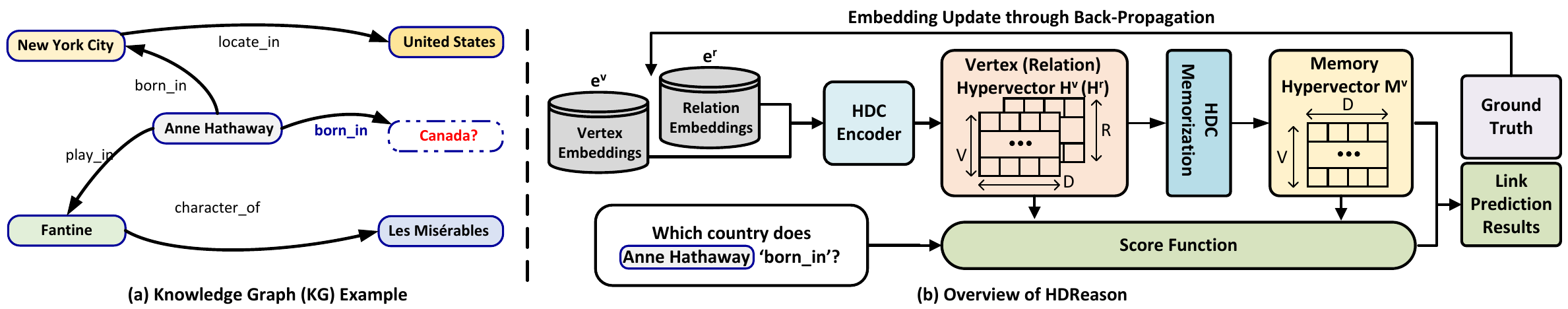}
  \caption{(a) KG example. (b) Overview of HDReason.}
  \vspace{-2mm}
  \label{Fig:HDReason_top}
\end{figure*}

\subsection{Knowledge Graph Link Prediction} \label{sec:KG_basic}

\begin{table}[t]
\caption{Features of HDReason in comparison with prior works.}
\label{tab:motivation}
\resizebox{.48\textwidth}{!}{%
\begin{tabular}{c|ccccc}
\toprule
Accelerator                             & GraphACT~\cite{zeng2020graphact} & HP-GNN~\cite{lin2022hp} & FlowGNN~\cite{10071015} & PyG~\cite{vashishth2019composition}* & \textbf{HDReason} \\ \midrule
Platfrom                                & FPGA     & FPGA   & FPGA    & GPU  & \textbf{FPGA}     \\ \midrule
Model Type                              & GNN      & GNN    & GNN     & GNN  & \textbf{HDC}      \\
Reasoning Support                       & No       & No     & No      & Yes  & \textbf{Yes}      \\
Edge Embedding                          & No       & No     & Yes     & Yes  & \textbf{Yes}      \\
\multicolumn{1}{l|}{Embedding Training} & No       & No     & No      & Yes  & \textbf{Yes}      \\
Interpretability                        & Low      & Low    & Low     & Low  & \textbf{High}     \\
Energy Efficiency                       & High     & High   & High    & Low  & \textbf{High}     \\
\multicolumn{1}{l|}{Computation Reuse}  & No       & No     & No      & No   & \textbf{Yes}      \\ \bottomrule
\end{tabular}}
\vspace{-3mm}
\end{table}

Figure~\ref{Fig:HDReason_top}(a) illustrates the basic concepts of KG and provides an example of the KGC task. A KG can be formally represented as a directed graph $G = \{(v, \,r, \, u)\:\vert\: u, v \in \xi, \:r \in R\}$, where $\xi$ is the set of entities (vertices) and $R$ is the set of relations~\cite{vashishth2019composition}. 
Each directed edge in a KG represents a factual statement and is defined as a fact triple $l = (v, \,r, \,u)$. As shown in Figure~\ref{Fig:HDReason_top}, (\texttt{Anne Hathaway}, \texttt{born}\_\texttt{in}, \texttt{New York City}) is a fact triple, which presents the statement \texttt{Anne Hathaway was born in the New York City}. 
Notice that relations are directed edges to distinguish the subject and object in factual statements.

In various KG-related reasoning applications, we focus on the KGC task due to its elevated computational demands and training challenges~\cite{chen2020review}. The KGC task is also referred to as KG link prediction. As a low-level logic reasoning~\cite{gao2021quatde}, KGC aims to infer nonexistent relations from available knowledge. An example task is to find a new triple like (\texttt{Anne Hathaway}, \texttt{born}\_\texttt{in}, ?) in the context of Figure~\ref{Fig:HDReason_top}(a).

Prior works generally handle two types of KGC: \textit{single direction reasoning} and \textit{double direction reasoning}. The former is limited to finding a possible connection from the source vertex to the destination vertex. And the latter requires finding both positive connection (v,r,?) and negative connection (?,r,u).

\subsection{Score Function}
The score function measures the distance between two nodes given its relation type~\cite{zheng2020dgl} in a KG link prediction task. For a subject $\vec{e^v_i}$ and the relation $\vec{e^r_k}$, if vertex $j$ is connected with vertex $i$ via relation $r_k$, the mapping function f is defined as:
\begin{equation} \label{eq:TransE0}
    \vec{e^v_j} \approx f(\vec{e^v_i}, \vec{e^r_k})
\end{equation}
Figure~\ref{Fig:HDReason_base}(d) provides an example of TransE~\cite{bordes2013translating}, a type of score function. TransE treats the relationship as a translation vector so that the embedded entities are connected by relation r with low error. The TransE score function is defined as:
\begin{equation} \label{eq:TransE1}
    score_j = N_x(\vec{e^v_i} + \vec{e^r_k} - \vec{e^v_j})
\end{equation}
Here $N_x$ stands for the norm function. 
$\vec{e^v_i}$, $\vec{e^r_k}$, and $\vec{e^v_j}$ can be also treated as subject vector, relation vector, and object vector. From a semantic point of view, equations~\ref{eq:TransE0} and \ref{eq:TransE1} stand for the subject being connected with the object via a specific relation. Given the query subject and relation, we use the TransE score function to calculate the score of every vertex in the graph. A larger score means that this vertex has a higher chance of being connected with the query subject vertex via the query relation. 

\subsection{KGC Challenge and Motivation}
In table~\ref{tab:motivation}, we summarize prior efforts that try to speed up GNN models. Historically, most GNN-based model accelerations leveraged GPUs built on the PyG framework~\cite{vashishth2019composition}. While GPUs are versatile computing platforms, they suffer from static data paths and reduced energy efficiency~\cite{lacey2016deep}. 
In contrast, FPGAs offer a balance between computational parallelism and energy consumption, making them suitable for enhancing GNN-based models. Nevertheless, it is challenging to apply existing FPGA-based GCN training accelerators directly to solve KGC tasks~\cite{zeng2020graphact,lin2022hp}. 
One major challenge is that KG datasets incorporate semantic data for both vertices and edges. The cutting-edge FPGA-based accelerator, FlowGNN~\cite{sarkar2023flowgnn}, does support these embeddings, but solely for GNN model inference. Furthermore, current FPGA accelerators typically focus on GNN propagation weight training, without supporting the crucial vertex and relation embedding training. Given the semantic richness of KG datasets, this training is central to the entire reasoning process. 
KGC training on GPU suffers from large-scale GPU memory usage for updating large-scale embeddings, extensive epochs for learning semantic information, and complexities in integrating accelerated score functions and graph models on FPGA platforms. A hardware-software approach is vital for an optimal FPGA framework tailored for KGC training.

\section{HDReason Model} \label{sec:KG_HDC}

\begin{table}[t]
\centering
\caption{HDReason Notation}
\vspace{-2mm}
\label{tab:HDReason_notation}
\resizebox{0.45\textwidth}{!}{%
\begin{tabular}{c|cc}
\toprule
\textbf{Notation} & \textbf{Definition}                                          & \textbf{Dimension}          \\ \midrule
$\vert V \vert$               & KG Vertex Size                               & $\sim$                      \\
$\vert R \vert$               & KG Relation Size                             & $\sim$                      \\
d                 & The original embedding dimension                          & $\sim$                      \\
D                 & The hyperspace dimension                                  & $\sim$                      \\
$\textbf{e}^\textbf{v}$                & The original space vertex embedding                       & $\vert V\vert\times d$                       \\
$\textbf{e}^\textbf{r}$                & \multicolumn{1}{l}{The original space relation embedding} & $\vert R\vert\times d$                       \\
$\textbf{H}^\textbf{B}$                & The encoding base hypervector matrix                      & $d\times D$                         \\
$\textbf{H}^\textbf{v}$                & The vertex hypervector matrix                             & $\vert V \vert\times D$                       \\
$\textbf{H}^\textbf{r}$                & The relation hypervector matrix                           & $\vert R \vert\times D$                       \\
$\textbf{M}^\textbf{v}$                & The vertex memory hypervector matrix                      & $\vert V \vert\times D$                       \\
$\textbf{M}_\textbf{q}^\textbf{v}$               & Query subject hypervector                                 & $1\times D$                         \\
$\textbf{M}_\textbf{q}^\textbf{r}$               & Query Relation  hypervector                               & $1\times D$                         \\
$N(\cdot)$               & The normalization function                                & $\sim$                      \\
$\textbf{P}$                 & The scores for each vertex                                & $1\times \vert V \vert$ \\ \bottomrule

\end{tabular}%
}
\vspace{-3mm}
\end{table}

In this section, we extend the original HDC graph learning model~\cite{nunes2022graphhd} to conduct KGC or link prediction. Figure~\ref{Fig:HDReason_top}(b) illustrates the reasoning process. Table~\ref{tab:HDReason_notation} lists the notations used in the \Design model. A KG denoted as \textbf{G}($\xi$, $R$) is consists of $|V|$ vertices and $|R|$ relations.

\subsection{HDReason Link Prediction}
The embedding vectors for all vertices comprise the embedding matrix $\textbf{e}^\textbf{v}$, with a dimension of $|V| \times d$. Similarly for the relations, we have an embedding matrix $\textbf{e}^\textbf{r}$ of size $|R| \times d$. To map all vertices and relations from the original embedding space into hyperspace, we have the matrix multiplication equation:
\begin{equation}
    \mathbf{H^v} = tanh(\mathbf{e^v}\mathbf{H^B})
\end{equation}
\begin{equation}
    \mathbf{H^r} = tanh(\mathbf{e^r}\mathbf{H^B})
\end{equation}
$\textbf{H}^\textbf{v}$ and $\textbf{H}^\textbf{r}$ are the encoded vertex and relation hypervector matrix with dimension $|V| \times D$ and $|R| \times D$ respectively.
 
After generating the vertex and relation hypervector matrix, we use an aggregation operation over the vertex and its neighbors to generate the corresponding memory hypervector as stated in equation~\ref{eq:memorization_base}. Given that each edge now also has a hyperspace mapping, it's essential to first execute binding operations between each vertex hypervector and its related relation hypervectors, ensuring the vertex data is correctly linked with its relevant relation information:
\begin{equation} \label{eq:memorize_kg}
    \vec{M^{v}_i} = \Sigma_{(j,r)\in N(i)} \Vec{H^v_j} \circ \Vec{H^r_r}
\end{equation}
, where $\circ$ represents element-wise Hadamard product. Equation~\ref{eq:memorize_kg} can also be rewritten in a matrix-to-matrix multiplication (M2MM) format:
\begin{equation} \label{eq:memorize_kg1}
    \mathbf{M^v} = \Sigma_{r \in R}((\mathbf{A^r}\mathbf{H^v}) \circ \mathbf{E^r})
\end{equation}
$\mathbf{A^r}$ is the adjacency matrix over relation r, with a dimension of $|V| \times |V|$. If there exists an entity ($v_i$, $r$, $v_j$) $\in$ $\xi$, then $A^r_{ij} = 1$, otherwise $A^{r}_{ij} = 0$. Matrix $\mathbf{E^r}$ with dimension $|V| \times D$ is the concatenation of relation embedding hypervectors across all vertices:
\begin{equation}
    \vec{E^r_i} = \vec{e_r} \quad i \in [0:|V|-1]
\end{equation}
After obtaining vertex memory hypervectors, we use a score function such as the backend decoder to perform link prediction tasks. This encoder-decoder structure is widely adopted by previous GNN-based works~\cite{schlichtkrull2018modeling,shang2019end,vashishth2019composition}. Here we choose to use TransE~\cite{bordes2013translating} as the score function. Assuming the input query is subject $\vec{e^v_i}$ and relation $\vec{e^r_k}$. To find the object vertex that is connected to subject vertex $v^i$ via relation $r^k$, we use the following vertex score function:
\begin{equation}
    \Vec{P} = Sigmoid(|M^v_i + H^r_k - \mathbf{M^v}|_1 + bias)
\end{equation}
The score vector P has a shape of $|V| \times 1$, with its element representing the possibility of a vertex connecting with vertex $v_i$ through relation $r$. In the equation above, the vectors $M^v_i$ and $H^r_k$ are first broadcasted to the same shape as the matrix $\mathbf{M^v}$. The L1 normalization is carried out only on one axis of the matrix, giving a vector form.

\subsection{HDReason Model Training}
On top of previous HDC graph learning works~\cite{poduval2022graphd,nunes2022graphhd,kang2022relhd}, we introduce the original space embedding to vertices and relations; we also add a score function to the HDC graph model. Now we can conduct backpropagation and train the original embedding model ($\textbf{e}^\textbf{v}$ and $\textbf{e}^\textbf{r}$):
\begin{equation} \label{eq:training_back}
    \nabla_{\mathbf{e^v}}\mathbf{L} = \frac{\partial \mathbf{L}}{\partial \mathbf{P}}\frac{\partial \mathbf{P}}{\partial \mathbf{M^v}}\frac{\partial \mathbf{M^v}}{\partial \mathbf{H^v}}\frac{\partial \mathbf{H^v}}{\partial \mathbf{e}^v}
\end{equation}
\begin{equation}
    \nabla_{\mathbf{e^r}}\mathbf{L} = \frac{\partial \mathbf{L}}{\partial \mathbf{P}}\frac{\partial \mathbf{P}}{\partial \mathbf{H^r}}\frac{\partial \mathbf{H^r}}{\partial \mathbf{e}^r}
\end{equation}
Here \textbf{L} is the training loss and \textbf{P} includes link prediction scores for vertices. The dimension of \textbf{P} and \textbf{L} is $1 \times |V|$. 

Although both \Design and previous GNN model~\cite{schlichtkrull2018modeling,vashishth2019composition,shang2019end} use front-end encoder (graph model) and backend score function structure, the base hypervector matrix remains fixed in hyperdimensional computation. This leads to more efficient learning since HDC only updates the original embedding of vertex and relation. 

\subsection{The Interpretability of HDReason} 

Compared to conventional GNN models~\cite{schlichtkrull2018modeling,vashishth2019composition,shang2019end}, HDC boasts enhanced memorization and information-retrieval proficiencies, allowing for the reconstruction of neighboring data for each vertex, as showcased in~\cite{poduval2022graphd}. As a result, the memorization hypervector ($\textbf{M}^\textbf{v}$) of each vertex symbolically represents its neighboring vertex's details. In contrast to the standard GNN model, \textbf{\Design offers superior interpretability and is more aligned with the principles of artificial general intelligence (AGI)} as indicated in~\cite{latapie2021metamodel}.

\section{FPGA Acceleration Design}~\label{sec:architecture_part}
\subsection{Architecture Overview}
\begin{figure}[t!]
  \centering
  \includegraphics[width=0.7\linewidth]{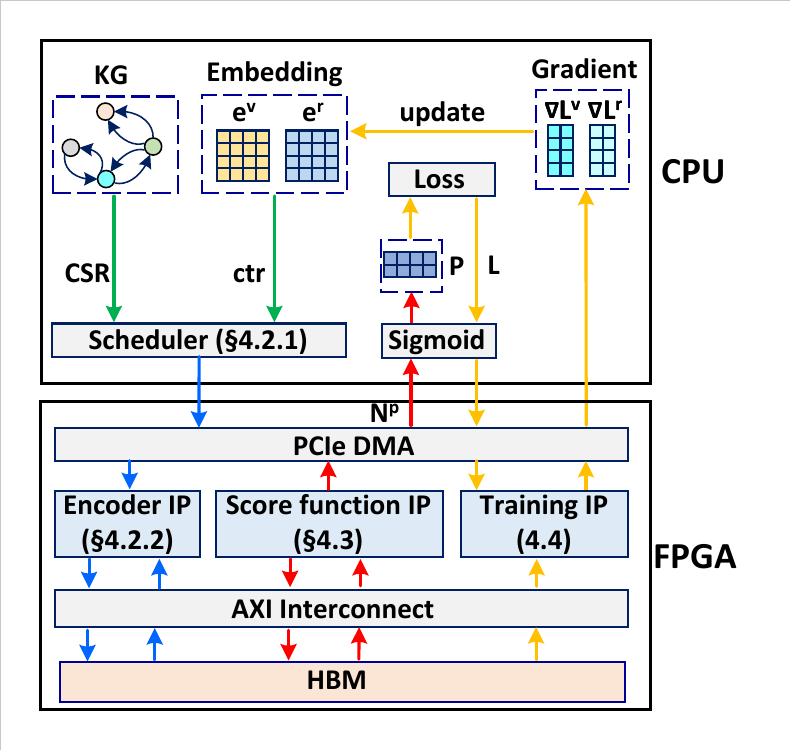}
  \vspace{-3mm}
  \caption{CPU-FPGA acceleration platform overview.}
  \small \textcolor{blue}{$\rightarrow$}: Encoder IP dataflow \textcolor{red}{$\rightarrow$}: Score function IP dataflow \textcolor{yellow}{$\rightarrow$}: Training IP dataflow
  \vspace{-3mm}
  \label{Fig:cpu-fpga}
\end{figure}

\begin{figure*}[t!]
  \centering
  \includegraphics[width=1\linewidth]{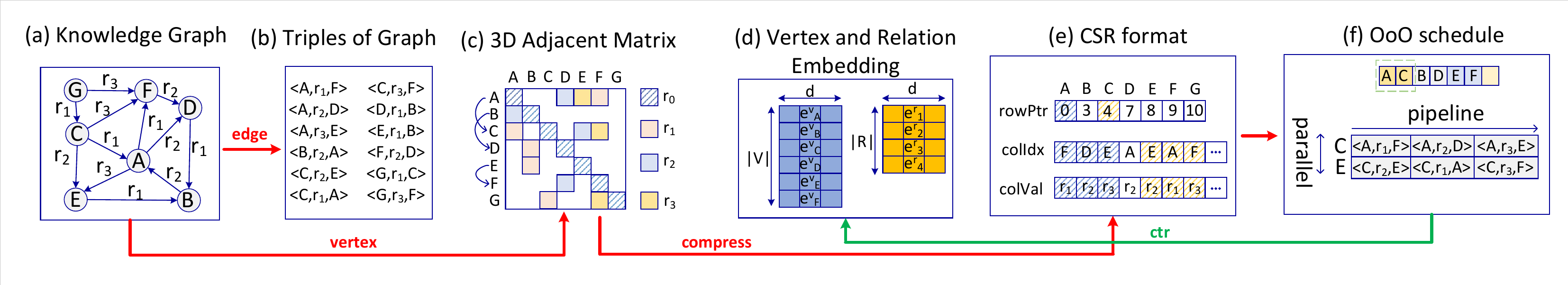}
  \caption{Balanced computation scheduling example. \textbf{CSR} means compressed sparse row and \textbf{OoO} means out of order.}
  \vspace{-2mm}
  \label{Fig:Scheduler_example}
\end{figure*}

This section introduces the architecture of \Design, tailored for a CPU-FPGA heterogeneous computing platform, as illustrated in Figure~\ref{Fig:cpu-fpga}. Our design is comprised of two parts: (1) the CPU component (host side) and (2) the FPGA component (kernel side), the latter one is where the original vertex and relation embeddings are managed and updated. The \textbf{CPU-based scheduler} is responsible for offloading vertex and relation data to the FPGA kernel. Within the FPGA kernel, the \textbf{Encoder IP} maps the vertex and relation embedding from the original space into the hyperspace and then performs the HDC memorization process. Every vertex's resulting memorization hypervector is stored in the high bandwidth memory (HBM). The \textbf{Score function IP} then calculates and dispatches the reasoning outcome back to the host CPU. During the backpropagation phase, the bulk of gradient computations are offloaded to the \textbf{Training IP}.

\subsection{Graph Memorization Acceleration} \label{sec:memorization}

\ifx{
\begin{algorithm}[b] 
\caption{Density-aware Scheduler}
\label{alg:OoO_scheduler}
\begin{algorithmic}[1]
\Require{rowPtr, colIdx, colVal, $N_c$, $e^v$, $T_G$}
\Statex
    \State {$H_G$ $\gets$ map$<$int, list$>$}
    \State {$M_G$ $\gets$ map$<$int, int$>$}
    \For{$i \gets 1$ to  $|V|$}                    
        \State {count = rowPtr[i+1] - rowPtr[i]}
        \State {$H_G$[count].add(i)}
        \State {$B_d$ = $f_0$($e^v$, $f_2$($H_G$[count], rowPtr, colIdx, $T_G$),$M_G$)}
        \State {$B_c$ = $f_1$($f_2$($H_G$[count], rowPtr, colIdx, $T_G$),$M_G$)}
        \State {call kernel($B_d$, $B_c$)}
        \State{$B_d$.clear(), $B_c$.clear()}
    \EndFor
    \State{L$\gets$ $f_2$($H_G$.list(), rowPtr, colIdx,$T_G$)}
    \For{i $\gets$ 1 to $len(L)$ by $N_c$ } 
        \For{$j \gets 1$ to $N_c$}
            \State {$B_d$ = $f_0$($e^v$,L[i:i+j],$M_G$)}
            \State {$B_c$ = $f_1$(L[i:i+j],$M_G$)}
            \State {call kernel($B_d$, $B_c$)}
            \State{$B_d$.clear(), $B_c$.clear()}
        \EndFor    
    \EndFor
\end{algorithmic}
\end{algorithm}

\begin{table}[t]
\centering
\caption{Notation of Scheduler Algorithm}
\vspace{-2mm}
\label{tab:scheduler_note}
\resizebox{0.45\textwidth}{!}{%
\begin{tabular}{c|c}
\toprule
\textbf{Notation}            & \textbf{Definition}                                                                                                                               \\ \midrule
$T_G$                        & Triples of KG                                                                                                                     \\
$H_G$                        & \multicolumn{1}{l}{Hash map with key as the number of neighbors and value as a list.}                                                          \\
$H_G{[}i{]}$                 & A list containing vertices whose neighbor size is i                                                                                            \\
$M_G$                        & \begin{tabular}[c]{@{}c@{}}Hash map with key as the vertex index and the value as \\ its encoded hypervector address on FPGA HBM.\end{tabular} \\
$B_d$                        & Buffer for input data to FPGA.                                                                                                                 \\
$B_c$                        & Buffer for input control signal to FPGA                                                                                                        \\
$f_0$                          & Adding vertex embedding into the buffer                                                                                                        \\
$f_1$                          & Adding control signal into the buffer                                                                                                          \\
$f_2$                          & Generating corresponding tiples including src, rel, and dst                                                                                    \\
L                           & A list including vertex index                                                                                                                  \\
\multicolumn{1}{l|}{kernel} & FPGA accelerator API                                                                                                                           \\ \bottomrule
\end{tabular}%
}
\vspace{-3mm}
\end{table}
}\fi

The KG memorization includes two parts. The first part is an out-of-order (OoO) scheduler running on CPU trying to balance different vertices' computations. The second part is a hardware acceleration IP conducting hypervector encoding and vertex neighbor aggregation.
\subsubsection{Density-aware Scheduler on CPU}
A scheduler running on the host CPU is used to balance the computation between different vertices on the FPGA. Figure~\ref{Fig:Scheduler_example} presents a scheduling example. A KG (Figure~\ref{Fig:Scheduler_example}(a)) generally could be represented in two different formats: \textbf{the triple format} used in Figure~\ref{Fig:Scheduler_example}(b), or \textbf{the adjacency matrix format} in Figure~\ref{Fig:Scheduler_example}(c). 
As mentioned in previous work~\cite{10071015}, if we keep both vertex and relation embedding for KGC (Figure~\ref{Fig:Scheduler_example}(d)), the adjacency matrix becomes a 3D matrix (Figure~\ref{Fig:Scheduler_example}(c)). However, existing hardware, neither GPU nor FPGA can support 3D matrix operations. Therefore modern GCN acceleration frameworks use scatter and reduce operations instead of sparse matrix-multplication (SpMM) to implement GNN~\cite{sarkar2023flowgnn}. However, directly applying scatter and gather operation kernel on the sparse graph will bring computation imbalance problem~\cite{song2022sextans}. \textbf{Our proposed scheduler can instantly balance the computation workloads of different vertices' aggregation operations, and thereby help sustain a high throughput on the kernel FPGA.} 
 
The key idea of our scheduler is to use triples representation and adjacency representation together during the offloading process. 
Suppose that the maximum vertex parallelism on the FPGA kernel is $\textbf{N}_\textbf{c}$. To balance the computation of each vertex, we store the index of vertices with the same neighbor size (Figure~\ref{Fig:Scheduler_example}(e)) into a list and offload these vertices to the FPGA when the length of the list equals $N_c$. In that way, each vertex's aggregation has the same level of computation complexity, as their neighbor sizes are identical (Figure~\ref{Fig:Scheduler_example}(f)).

Beyond ensuring balanced computations across vertices, the scheduler's pivotal role is to minimize redundant encoding operations. Typically, HDC models begin by encoding all data into hypervectors before making any predictions. Yet, in the context of KGC, \textbf{the sheer volume of triples often surpasses the count of vertices and relations}. Thus, using the same encoding-prediction paradigm may \textbf{incur redundant hypervector encoding operations}. 
To counter this inefficiency, we opt to store already encoded vertex hypervectors directly in the FPGA's memory, like HBM, while executing HDC encoding only for unencoded vertices.
On the CPU side, we maintain a HashMap that associates a vertex index (as the key) with its respective hypervector address (as the value)in the FPGA memory. 
If the vertex is not encoded, the scheduler will add the vertex's original embedding vector into the data buffer ($\textbf{B}_\textbf{d}$), otherwise, the corresponding FPGA memory address is added ($\textbf{f}_\textbf{1}$). Meanwhile, a control signal ($\textbf{f}_\textbf{2}$) indicating whether this vertex has been encoded and which vertex it is connected to is added to the control signal buffer ($\textbf{B}_\textbf{c}$). After filling the buffers, the CPU offloads the HDC computation into the FPGA kernel.

\begin{figure*}[t!]
  \centering
  \includegraphics[width=1\linewidth]{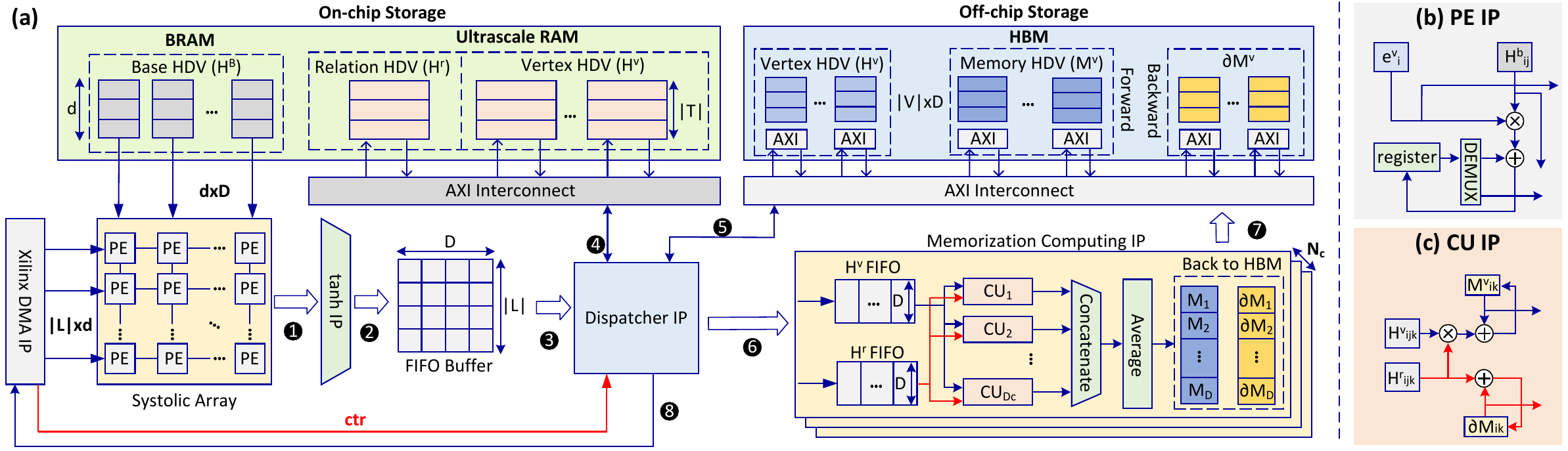}
  \caption{The encoder architecture design. \textbf{Systolic Array} encode embedding vector from normal space into hyperspace. \textbf{Dispatcher IP} dynamically load encoded hypervectors from off-chip memory into on-chip memory. \textbf{Memorization Computing IP} simultaneously execute the forward memorization and backward gradient computation}
  \vspace{-3mm}
  \label{Fig:encoder_IP}
\end{figure*}

\subsubsection{Encoder IP Architecture Design} 

Figure~\ref{Fig:encoder_IP} (a) shows the encoder IP architecture design. In every epoch, the host CPU offloads vertex embedding into the kernel FPGA via direct memory access (DMA) IP. The hyperdimensional encoding includes two steps. In the first step, each vertex embedding ($1 \times d$) is multiplied with the base hypervector matrix via a systolic array IP ($\invcircledast{1}$).
Figure~\ref{Fig:encoder_IP} presents the microarchitecture of each PE IP. Here we also pipeline the vertex-to-vertex computation into $|L|$ stages, where $|L|$ is the number of vertices that have not been encoded. The value of $|L|$ varies across different FPGA kernel calls. Subsequently, each encoded hypervector is passed through kernel function (here we choose \textbf{tanh}) ($\invcircledast{2}$). 
The encoded vertex hypervectors ($\textbf{H}_\textbf{v}$) are then pushed into a FIFO awaiting the \textbf{Dispatcher IP}'s processing ($\invcircledast{3}$).

To improve the HDC model's training throughput, it is important to \textbf{reuse encoded hypervectors}. In our design, the \textbf{Dispatcher IP} plays a pivotal role in achieving this goal. The Dispatcher IP in Figure~\ref{Fig:encoder_IP}(a) has three functionalities. \textbf{The first} is writing the newly encoded vertex hypervectors into the FPGA memory (e.g. HBM in Figure~\ref{Fig:encoder_IP}(a)) and returning the assigned address to the host CPU ($\invcircledast{8}$). \textbf{The second} is dispatching each vertex and its corresponding neighbor hypervectors and relation hypervectors into the Memorization Computing IP ($\invcircledast{6}$). \textbf{The third} is managing the on-chip memory storage (such as UltraRAM) and performing replacement policy when necessary ($\invcircledast{4}$ and $\invcircledast{5}$). 
Ideally we prefer to store all encoded hypervector on-chip to avoid redundant computation. However, the vertex size of the graph dataset is usually too large to store all vertex hypervectors on-chip. Therefore, we need a mechanism to efficient manage on-chip hypervectors like cache replacement mechanism~\cite{hmmu, openmem, softhint}. The strategy that we choose in Figure~\ref{Fig:encoder_IP} is to store all the relation embedding hypervectors ($\textbf{H}_\textbf{r}$) and only part of the vertex hypervectors. 
A HashTable is implemented with content-addressable memory (CAM) inside the Dispatcher IP. The key of the HashTable is the vertex id and the value is its corresponding hypervector address on Ultarscale RAM. If the encoded vertex id already resides in CAM, the hypervector will be loaded into the $H^v$ FIFO, while the corresponding relation hypervectors will also be loaded into the $H^r$ FIFO. If there is no room for a vertex hypervector in the on-chip UltraRAM, the Dispatcher IP will choose a victim hypervectors inside UltraRAM and replace it with the target vertex hypervectors from HBM. Here we choose the classic replacement algorithm such as last recent use (LRU), last frequently use (LFU), and random replacement policy~\cite{lee1999existence}.       

\begin{figure*}[t!]
  \centering
  \includegraphics[width=0.8\linewidth]{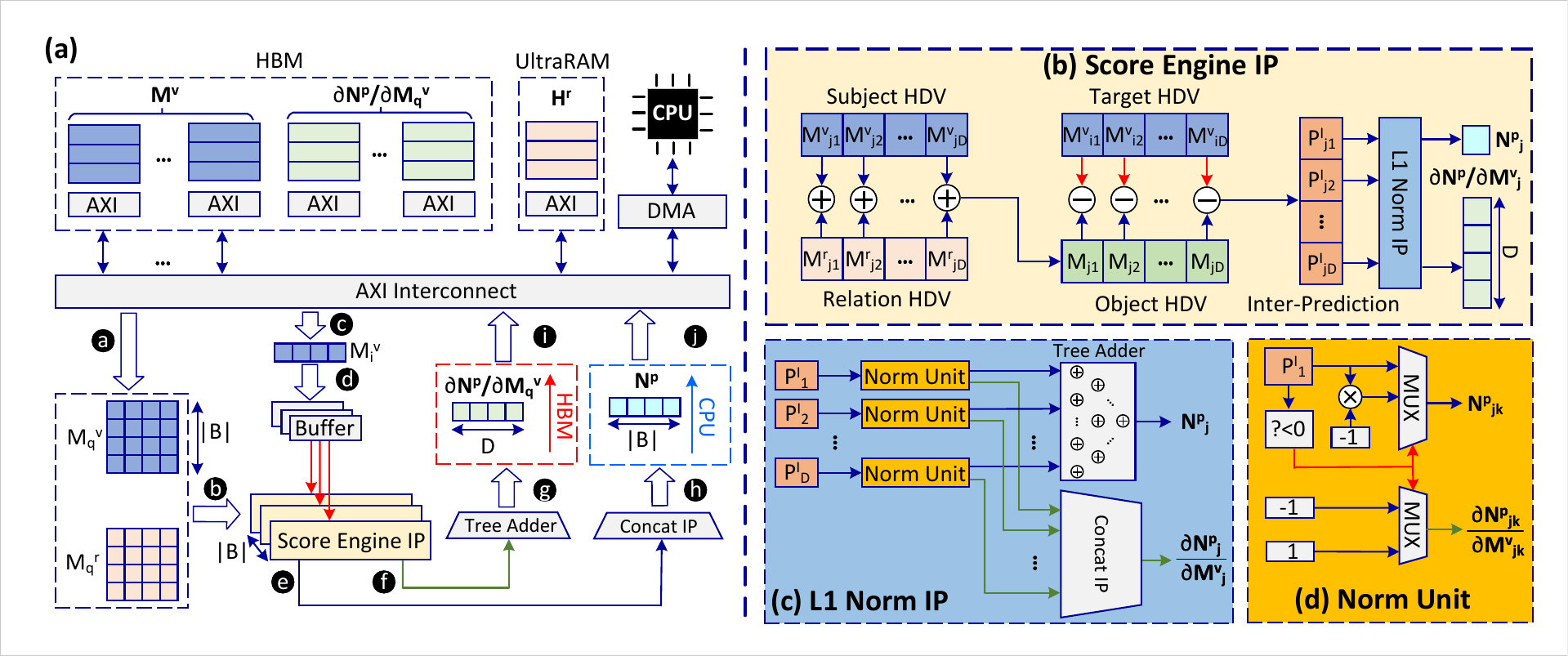}
  \vspace{-3mm}
  \caption{The score function architecture design.}
  \vspace{-3mm}
  \label{Fig:score_function_ip}
\end{figure*}

There are $N_c$ Memorization Computing IP on-chip that can perform neighbor aggregation operation for $N_c$ vertices concurrently. We pipeline the aggregation operation of each vertex's neighbors. The binding operation between the vertex hypervector and relation hypervector is parallelized among computing units (CU), whose microarchitecture is shown in Figure~\ref{Fig:encoder_IP}(c). Please note that we also need to compute the gradient of memory hypervector $\frac{\partial M^v}{\partial H^v}$:
\begin{equation} \label{eq:gradient_M}
    \frac{\partial \mathbf{M^v}}{\partial \mathbf{H^v}} =  \Sigma_{r \in R}\mathbf{A_r}\mathbf{E^r}
\end{equation}
Here matrix $A_r$ is a $|V| \times |V|$ matrix and is the same in equation~\ref{eq:memorize_kg1}. The gradient of each vertex's memory hypervector ($M^v$) is the sum of its connected edges' relation hypervectors. \textbf{Therefore, the CUs can execute the forward memorization and backward gradient computation together.} 

\subsection{Score Function Acceleration} \label{sec:score_function}
After generating memory hypervectors ($\textbf{M}^\textbf{v}$), the next step is to perform reasoning tasks over specific triples. For each inference or training epoch, suppose the batch size is $\vert B \vert$. As is shown in Figure~\ref{Fig:score_function_ip}.(a), at the initial stages ($\invcircledast{a}$ and $\invcircledast{b}$), we load query vertex hypervector matrix ($\textbf{M}_\textbf{q}^\textbf{v}$) and query relation hypervector matrix ($\textbf{M}_\textbf{q}^\textbf{r}$) onto the on-chip buffer from FPGA storage (HBM). As our goal is to predict whether each triple exists or not, we pipeline the score function calculation process and only load one memory hypervector at a time ($\invcircledast{c}$), instead of loading the entire memory hypervector ($M^v$). In Figure~\ref{Fig:score_function_ip}, we assume that the Score function IP is evaluating \textbf{vertex i} over the query batch. Since the query batch contains $|B|$ subject and relation combinations, we replicate the loaded memory hypervector into $|B|$ on-chip buffers ($\invcircledast{d}$). Next, the query hypervectors ($M_q^v$ and $M_q^r$) and the memory hypervectors ($M^v_i$) are loaded into $|B|$ Score Engine units, with each unit calculating the corresponding prediction score for each query.

Figure~\ref{Fig:score_function_ip}(b) presents a single Score Engine IP's microarchitecture design. Suppose the illustrated Score Engine unit in Figure~\ref{Fig:score_function_ip}(b) has an index of $j$, then it calculates $j^{th}$ vertex of the batch's score over vertex $i$. The \textbf{first step} is to perform hypervector addition over subject HDV (query vertex HDV $M^v_j$) and relation HDV ($M^r_j$) to obtain the object HDV ($M^o_j$). The \textbf{second step} is to calculate the hypervector distance between $M^o_j$ and $M^v_i$. The intermediate prediction result($P^I_j$) is loaded into the L1 Norm IP for L1 normalization over the hyperspace, resulting in the normalized prediction ($N^p_j$) and its gradient with respect to the queried vertex and relation hypervector.
Notice that these two gradients are actually equal in value since they are added together in the score function: $\frac{\partial N^p_j}{\partial M^v_j} = \frac{\partial N^p_j}{\partial M^r_j}$.
As mentioned in section~\ref{sec:memorization}, to reduce the computation cost of back-propagation and avoid unnecessary data movement, inside the L1 Norm IP, we perform gradient computation of $\frac{\partial N^p_j}{\partial M^v_j}$ during the forward propagation, as shown in Figure~\ref{Fig:score_function_ip}(c) and Figure~\ref{Fig:score_function_ip}(d). Inside L1 Norm IP, there are $D$ \textbf{Norm Units} in total. Each of them extracts the absolute values and signs of elements of the input hypervector. The results are then fed into Tree Adder IP and a concatenate IP (Concat IP) to generate norm prediction $N^p_j$ and $\frac{\partial N^p_j}{\partial M^v_j}$ respectively.

\textbf{Until now, after the execution of Score Engine IP, we have each batch member's both forward path and backward path results.} Next we pass the forward path result ($N^p_j$) and backward gradient result ($\frac{\partial N^p_j}{\partial M^v_j}$) to a Concat IP ($\invcircledast{e}$) and Tree Adder IP ($\invcircledast{f}$) respectively. To generate the final result, we need to concatenate each batch member's L1 norm result ($\invcircledast{g}$) and accumulate all batch member's gradient hypervectors ($\invcircledast{h}$). In the end, we will pass the forward path result ($N^p$) back to the CPU for further processing, such as the Sigmoid function ($\invcircledast{i}$). The gradient result ($\frac{\partial N^p_j}{\partial M^v_j}$) instead, will be stored inside the FPGA storage (such as HBM) for later training back-propagation usage ($\invcircledast{j}$).

\subsection{Symbolic Training Acceleration} \label{sec:training}
\begin{figure}[t!]
  \centering
  \includegraphics[width=1\linewidth]{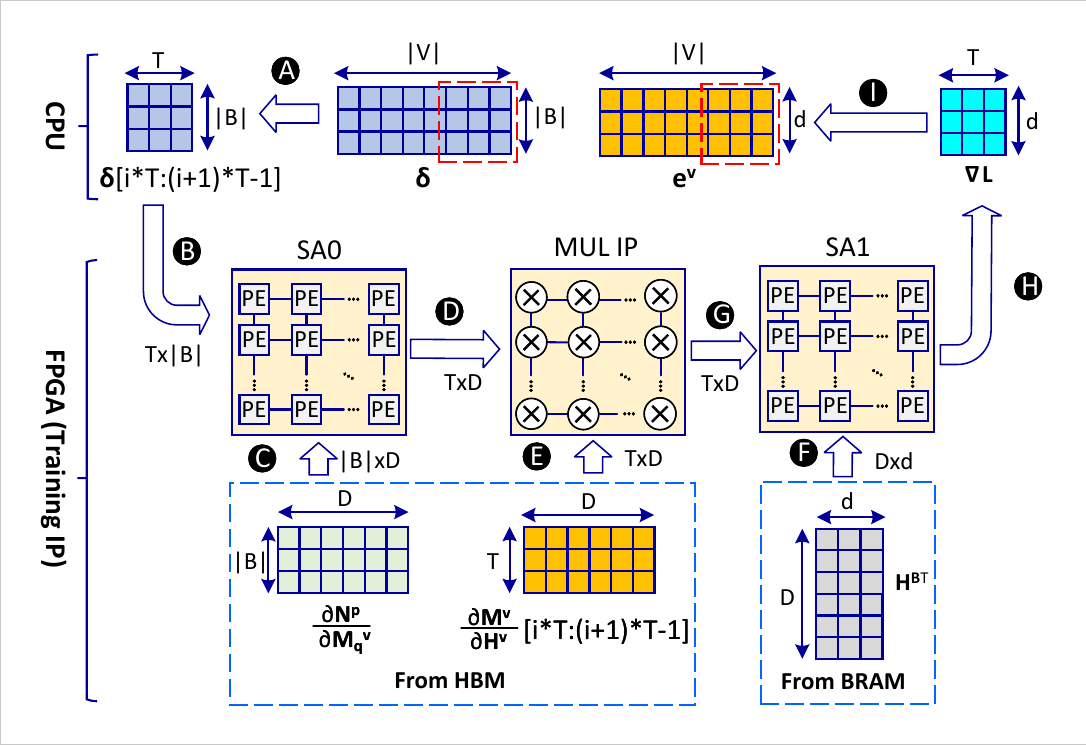}
  \vspace{-3mm}
  \caption{The vertex embedding training computing training process. Partial gradient result is prestored inside HBM.}
  \label{Fig:training_IP}
\end{figure}

Figure~\ref{Fig:training_IP} presents the vertex embedding training computing flow. For the readability of the figure, we omit the detail of relation embedding training as it is similar to Figure~\ref{Fig:training_IP}. For illustration convenience, we rewrite the backpropagation equation:
\begin{equation}
    \nabla_{\mathbf{e^v}}\mathbf{L} = \frac{\partial \mathbf{L}}{\partial \mathbf{N}}\frac{\partial \mathbf{N}}{\partial \mathbf{N^p}}\frac{\partial \mathbf{N^p}}{\partial \mathbf{M^v}}\frac{\partial \mathbf{M^v}}{\partial \mathbf{H^v}} \frac{\partial \mathbf{H^v}}{\partial \mathbf{e}^v}
\end{equation}
We first rewrite the first two terms as:
\begin{equation} \label{eq:CPU_gradient}
    \mathbf{\delta^v} = \frac{\partial \mathbf{L}}{\partial \mathbf{N}}\frac{\partial \mathbf{N}}{\partial \mathbf{N^p}}
\end{equation}
We will calculate each batch member's $\delta^v$ on CPU and concatenate them together. The generated new matrix is named $\mathbf{\delta}$ whose dimension is $|B| \times |V|$ where $|B|$ is the batch size and $|V|$ is the vertex size of the graph. Since graph size $|V|$ could be very large (over 10K), we cut $\delta$ into several chunks ($\invcircledast{A}$). The size of each chunk is $|B| \times T$. In Figure~\ref{Fig:training_IP}, we suppose the chunk index is i. So this chunk could be represented as $\delta [i*T:(i+1)*T-1]$. We also pipeline the computation between different chunks which means we will update T vertex embedding at each period.

\begin{figure*}[t!]
  \centering
  \includegraphics[width=1\linewidth]{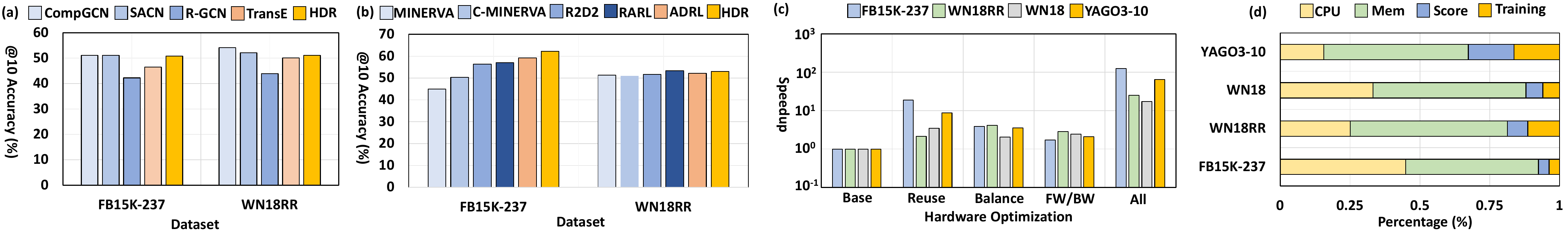}
  \vspace{-6mm}
  \caption{(a) Double direction reasoning accuracy comparison. HDR means HDReason with $D=256$. (b) Single direction reasoning accuracy. (c) Hardware optimization effects. (d) Heterogeneous computing platform execution time breakdown. }
  \vspace{-3mm}
  \label{Fig:accuray_and_breakdown}
\end{figure*}

Next we pass the $\delta [i*T:(i+1)*T-1]$ to the kernel FPGA ($\invcircledast{B}$). \textbf{In section~\ref{sec:memorization} and~\ref{sec:score_function}, we already compute the gradient result of $\frac{\partial N^p}{\partial M^v}$ and $\frac{\partial M^v}{\partial H^v}$ during the forward propagation} and stored them inside the FPGA storage (such as HBM). Also the gradient of $\frac{\partial \mathbf{H^v}}{\partial \mathbf{e}^v}$ is equal to the transpose of base HDV ($\textbf{H}^\textbf{B}$). We only need to load precomputed matrices into the on-chip buffer ($\invcircledast{C}$, $\invcircledast{E}$, and $\invcircledast{G}$) and use two systolic arrays (SA) IP and one element-wise multiplication (MUL) IP to perform the actual computation ($\invcircledast{B}$, $\invcircledast{D}$, and $\invcircledast{F}$). After the computation of systolic array IP 2 (SA2), we will get T vertex's gradient result $\nabla$L which will be passed back to the host CPU ($\invcircledast{H}$). The last step is updating the T vertex embedding model ($\invcircledast{I}$) which is also the last stage of the whole pipeline computation. 

\section{Experiments} \label{sec:res}
\subsection{Experimental Setup}

\begin{table}[t]
\centering
\caption{KGC datasets statistics.}
\label{tab:dataset}
\resizebox{0.45\textwidth}{!}{%
\begin{tabular}{ccccccc}
\toprule
\textbf{Dataset} & \textbf{Entities} & \textbf{Relations} & \textbf{Train} & \textbf{Valid} & \textbf{Test} & \textbf{Avg. degree} \\ 
\midrule
\textbf{FB15K-237} & 14541 & 237 & 272115 & 17535 & 20466 & 18.71 \\
\textbf{WN18RR} & 40943 & 11 & 86835 & 3034 & 3134 & 2.12 \\
\textbf{WN18} & 40943 & 18 & 141442 & 5000 & 5000 & 3.45 \\
\textbf{YAGO3-10} & 123182 & 37 & 1079040 & 5000 & 5000 & 8.76 \\
\bottomrule
\end{tabular}%
}
\vspace{-3mm}
\end{table}

We benchmark \Design's KGC accuracy over two standard datasets: FB15K-237~\cite{toutanova2015observed} and WN18RR~\cite{dettmers2018convolutional} which are widely used for graph-related tasks. For \Design's hardware acceleration performance, we also added two large-scale datasets: WN18~\cite{miller1995wordnet} and YAGO3-10~\cite{mahdisoltani2014yago3}. Table~\ref{tab:dataset} lists the properties of these four KG datasets.
Our platform comprises an Intel i9-12900KF CPU as the host, along with a Xilinx Alveo U280 (or U50 depending on the configuration) card to implement the FPGA kernels.
We implement the kernel accelerator using SystemVerilog and synthesize it using Xilinx Vivado. The communication between the host CPU and kernel FPGA is supported by the Xilinx Vitis platform via PCIe direct memory access (PCIe DMA)~\cite{kathail2020xilinx}.    

\subsection{HDReason Reasoning Results}

\begin{figure}[t!]
  \centering
  \includegraphics[width=1\linewidth]{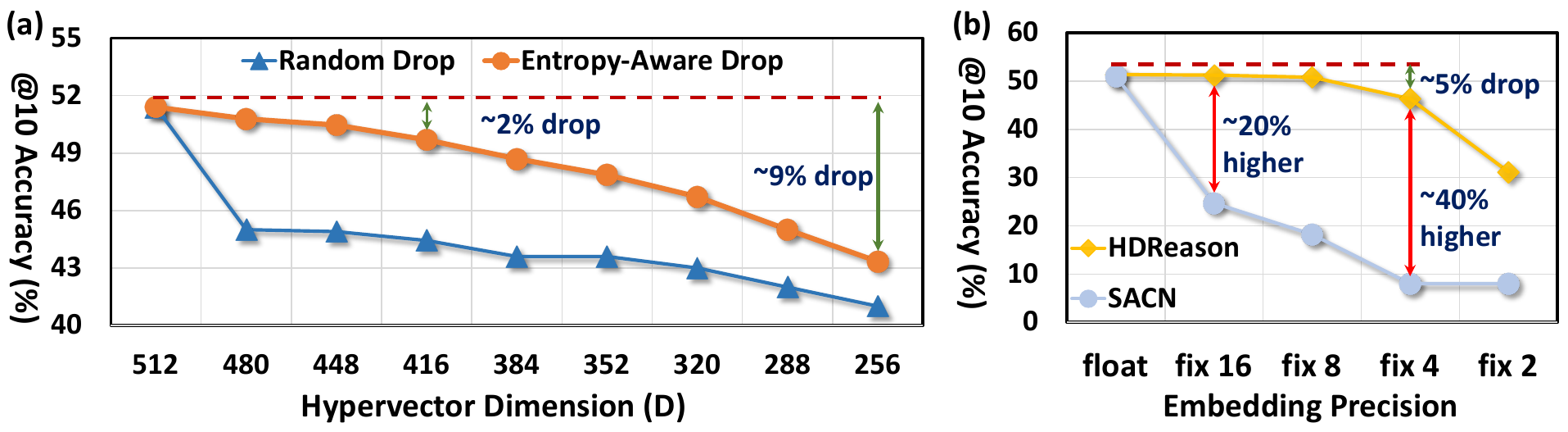}
  \vspace{-4mm}
  \caption{(a) HDC dimension drop. (b) Quantization effects comparison between HDC and GNN-based model. Here fix-N means fixed-point number with N bits.}
  \label{Fig:motivation_res}
\end{figure}

\begin{table}[t]
\caption{Knowledge Graph Model Comparison.}
\label{tab:model_compare}
\resizebox{0.45\textwidth}{!}{%
\begin{tabular}{c|ccc|cc}
\toprule
Model                              & CompGCN~\cite{vashishth2019composition}    & SACN~\cite{shang2019end}          & R-GCN~\cite{schlichtkrull2018modeling}       & TransE~\cite{bordes2013translating}        & \textbf{HDR}       \\ \midrule
d                                  & 100        & 100           & 100         & 150           & 128                \\
D                                  & 150        & 100           & 100         & $\sim$        & 256                \\
layer                              & 2          & 1             & 2           & $\sim$        & $\sim$             \\
$f_{score}$                           & TransE     & Conv-TranE    & DistMult~\cite{yang2015embedding}    & $\sim$        & TransE             \\ \midrule
\multicolumn{1}{l|}{Training Part} & \multicolumn{3}{c|}{embedding and weight} & \multicolumn{2}{c}{\textbf{only embedding}} \\ \bottomrule
\end{tabular}%
}
\vspace{-3mm}
\end{table}

Figure~\ref{Fig:accuray_and_breakdown}.(a) and Figure~\ref{Fig:accuray_and_breakdown}.(b) present the \Design's reasoning accuracy over FB15K-237 and WN18RR. For double direction reasoning, we compare our HDC-based model with other state-of-the-art graph models including TransE~\cite{bordes2013translating}, {R-GCN}~\cite{schlichtkrull2018modeling}, SACN~\cite{shang2019end}, and Comp-GCN~\cite{vashishth2019composition}. For single direction reasoning, we compare our HDC-based model with other state-of-the-art reinforcement learning (RL) models including MINERVA~\cite{das2017go}, C-MINERVA~\cite{stoica2020contextual}, R2D2~\cite{hildebrandt2020reasoning}, RARL~\cite{wu2021findings}, and ADRL~\cite{wang2020adrl}. 
Table~\ref{tab:model_compare} summarizes critical model parameters including original vertex embedding dimension d and final embedding dimension D (after encoding and aggregation). Figure~\ref{Fig:accuray_and_breakdown}.(a) shows that \Design achieves relatively high reasoning accuracy compared to state-of-the-art GCNs such as CompGCN and SACN. Besides, \Design only needs to train vertex and relation embedding, which simplifies the model training computation complexity. 
     
Figure~\ref{Fig:motivation_res}.(a) illustrates the HDR model's high robustness during KGC tasks. After encoding and memorization, we attempt to drop some hypervector dimensions before entering the score function. HDC model achieves significant robustness when dropping dimensions with low entropy. This attribute is also reported by previous HDC work~\cite{zou2021scalable}. This characteristic allows the HDC model to be deployed on resource-constrained platforms by maintaining only a portion of the model's hypervector weight. On the other hand, Figure~\ref{Fig:motivation_res}.(b) demonstrates that, when compared to a traditional GNN-based model (specifically SACN), HDR supports aggressive low-bit quantization. Here we try to quantize HDR and SACN into different fixed-point numbers using QPyTorch~\cite{zhang2019qpytorch}. When quantizing the HDR model from floating-point numbers to 4-bit fixed-point numbers, it experiences only a 5\% drop in accuracy, whereas the GNN-based model experiences a 45\% accuracy drop. This indicates that HDR is well-suited for low-bit precision computation platforms, such as FPGA.

\subsection{FPGA Resource Utilization}

\begin{table}[t]
\caption{Resource usage on Xilinx Alveo U50 FPGA.}
\label{tab:FPGA_resource}
\resizebox{0.45\textwidth}{!}{%
\begin{tabular}{c|ccccc}
\toprule
\multicolumn{1}{l|}{}                           & \textbf{LUT}               & \textbf{FF}                & \textbf{BRAM} & \multicolumn{1}{l}{\textbf{UltraRAM}} & \textbf{DSP} \\ \midrule
\textbf{Available}                              & 872K                       & \multicolumn{1}{l}{1743K}  & 1344          & 640                               & 5952         \\ \midrule
\textbf{Encoder IP}                             & \multicolumn{1}{l}{281.6K} & 152K                       & 184           & 135                               & 1024         \\
\multicolumn{1}{l|}{\textbf{Score Function IP}} & \multicolumn{1}{l}{238.9K} & \multicolumn{1}{l}{417.1K} & 0             & 0                                 & 0            \\
\textbf{Training IP}                            & 7.6K                       & 8.7K                       & 0             & 0                                 & 1536         \\
\textbf{HBM}                                    & 544                        & 437                        & 2             & 0                                 & 0            \\
\textbf{Others}                                 & 91.2K                      & 88.9K                      & 124           & 0                                 & 0            \\ \midrule
\textbf{Total}                                  & 620K                       & 667.2K                     & 310           & 135                               & 2560         \\ 
\textbf{Percentage}                             & 71.1\%                     & 38.2\%                     & 23.1\%        & 21\%                              & 43\%         \\ \midrule
\textbf{Freqeuency}                             & \multicolumn{5}{c}{200 MHz}                                                                                                \\
\textbf{Power}                                  & \multicolumn{5}{c}{36.1 W}                                                                                                 \\ \midrule
\end{tabular}%
}
\vspace{-3mm}
\end{table}

\begin{table*}[t]
\caption{HDReason's single batch training absolute results of latency and energy consumption on FPGA and GPU.}
\label{tab:FPGA_vs_GPU}
\resizebox{\textwidth}{!}{%
\begin{tabular}{c|cccc|c|cccc}
\toprule
                      & \multicolumn{4}{c|}{\textbf{Xilinx Alveo U50}}                    &                              & \multicolumn{4}{c}{\textbf{NVIDIA RTX 3090}}             \\
                      & \multicolumn{4}{c|}{Tech: UltraScale+ 14 nm FinFET; Freq: 200MHz} &                              & \multicolumn{4}{c}{Tech: Ampere TSMC 8nm; Freq: 1.66GHz} \\ 
                      & \multicolumn{4}{c|}{Memory Type: HBM2; Memory Bandwidth: 460GB/s} &                              & \multicolumn{4}{c}{Memory Type: GDDR6 Memory Bandwidth: 936.2 GB/s} \\ \midrule
                      & FB15K-237         & WN18RR        & WN18         & YAGO3-10       &                              & FB15K-237      & WN18RR      & WN18       & YAGO3-10*     \\ \midrule
\textbf{Latency (ms)} & 6.21              & 9.01          & 10.03        & 30.31          & \textbf{Latency (ms)}        & 60.01          & 91.01       & 93.62      & 219.6       \\
\textbf{Energy (J)}   & 0.21              & 0.29          & 0.31         & 0.93           & \textbf{Energy (J)} & 20.88          & 30.48       & 30.89      & 65.31       \\
\textbf{Memory (MB)}  & 33                & 84            & 86           & 245            & \textbf{Memory (MB)}         & 9608           & 23360       & 18690      & 22498       \\ \bottomrule
\end{tabular}%
}
\vspace{-3mm}
\end{table*}

Table~\ref{tab:FPGA_resource} presents the FPGA resource utilization of \Design. The vertex initial embedding dimension (\textbf{d}) is 96 and the encoded hypervector dimension (\textbf{D}) is 256. The training batch size is 128 and pipeline chunk size \textbf{T} is 32. We use 8 HBM pseudo channels (PCs) and AXI with 256-bit data width: 4 PCs are used to store vertex ($\textbf{H}^\textbf{v}$) and memorization hypervectors ($\textbf{M}^\textbf{v}$); while another 4 PCs hold gradients including $\frac{\partial N^p}{\partial M^v}$ and $\frac{\partial M^v}{\partial H^v}$. We save relation hypervectors ($\textbf{H}^\textbf{r}$) in the on-chip UltraRAM. 
The "Other" section in Table~\ref{tab:FPGA_resource} includes all the Xilinx IP resource usage such as AXI Interconnect IP and PCIe DMA~\cite{feist2012vivado}. 

\subsection{HDreason Acceleration on FPGA}
Table~\ref{tab:FPGA_vs_GPU} tabulates the execution time, energy, and memory usage of \Design on FPGA and GPU. Our GPU implementation uses PyTorch~\cite{paszke2019pytorch} and PyG~\cite{fey2019fast}. For a fair comparison, we specify both FPGA and GPU's training batch size as 128 except YAGO3-10 on GPU. We collect the FPGA power using Xilinx Power Estimator (XPE) and GPU power using NVIDIA Management Library (NVML). When testing \Design on RTX 3090 using the YAGO3-10 dataset, due to limited GPU memory (24 GB) and large-scale graph vertex size (over 120K vertices), we decrease the training batch size from 128 to 32.
As for speedup, our accelerator shows, on average, over \textbf{9$\times$} improvement compared to GPU even Alveo U50 adopts less advanced fabrication technology compared to RTX 3090 (14 nm FinFET on FPGA vs 8 nm TSMC on GPU).
\textbf{Our FPGA-based accelerator shows notable advantages over GPU in terms of speed, energy, and memory efficiency.}  
This large advantage comes from our hardware optimization shown in Figure~\ref{Fig:accuray_and_breakdown}.(c), including reusing encoded hypervectors, a density-aware scheduler to balance the computation, and computing backward gradients in the forward path. 
For energy efficiency, our design shows over \textbf{95$\times$} improvement compared to RTX 3090. For memory efficiency, as we cut the large vertex embedding gradient computation into small chunks and pipeline computation between different chunks, we avoid using unnecessary intermediate memory on the FPGA. Therefore, our FPGA accelerator supports a high batch size on large-scale graph datasets (like YAGO3-10).            

Figure~\ref{Fig:accuray_and_breakdown}.(d) shows the execution time breakdown during each single batch training. Here, CPU time includes data communication time between host CPU and kernel FPGA, CPU gradient computation (equation~\ref{eq:CPU_gradient}), and vertex and relation embedding update. Since we compute partial backward gradient in the forward path, in Figure~\ref{Fig:accuray_and_breakdown}.(d), actual training computation (Training) only takes a very small part of the overall training process. The memorization (Mem) execution time instead takes over 50\% total execution time. In section~\ref{sec:parameter_tune}, we will discuss how to decrease the memorization execution time by tuning the accelerator's parameters.

\begin{figure}[t!]
  \centering
  \includegraphics[width=1\linewidth]{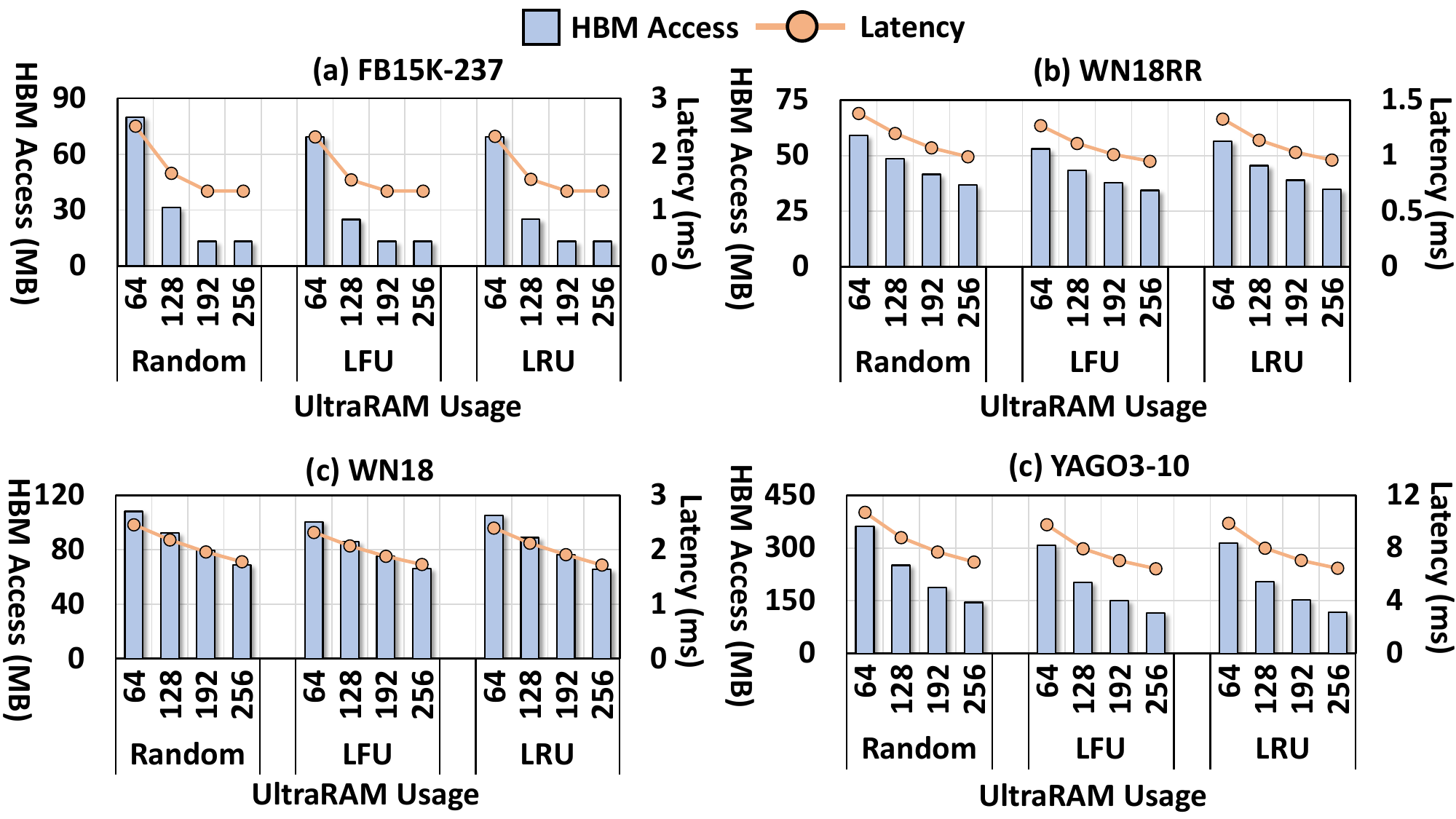}
  \caption{Replacement policy and on-chip UltraRAM usage effects to memorization speed.}
  \vspace{-5mm}
  \label{Fig:res_replace}
\end{figure}

\subsection{HDReason Memorization Improvement} \label{sec:parameter_tune}

\begin{figure*}[t]
  \centering
  \includegraphics[width=1\linewidth]{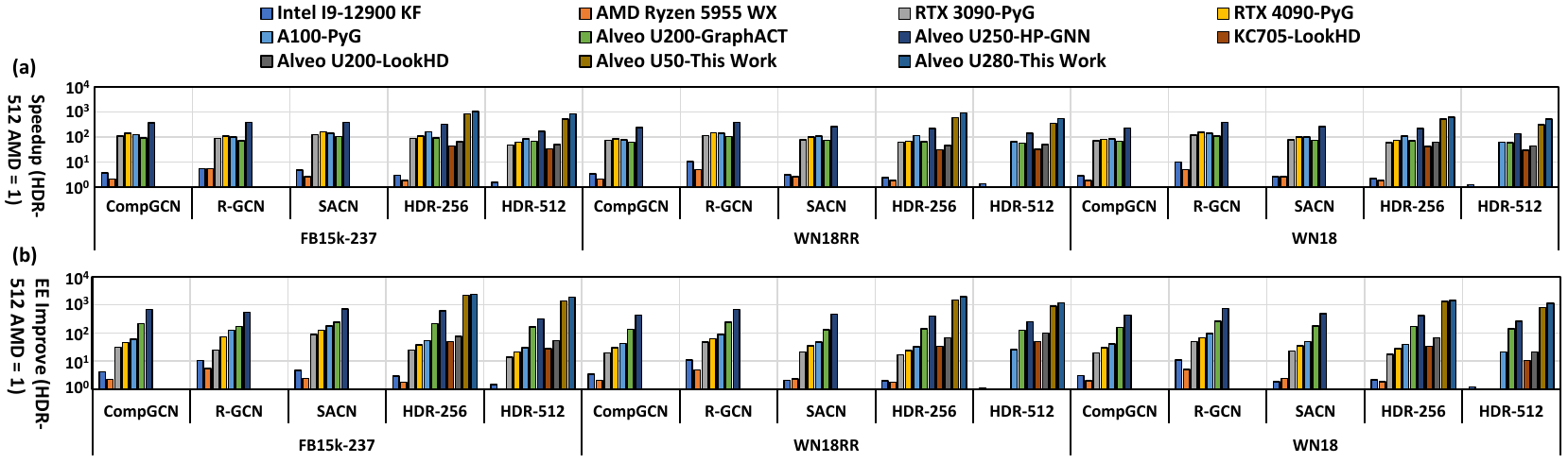}
  \vspace{-6mm}
  \caption{Cross models and cross platforms comparison of end to end model training. The batch size of all models and platforms is 128. Here EE Improve represents Energy Efficiency Improve.}
  \vspace{-3mm}
  \label{Fig:speedup_compare}
\end{figure*}

In this section, we discuss the effects of vertex replacement policy and on-chip storage usage over memorization speed. Compared to previous GCN acceleration works, our accelerator reuses encoded hypervectors such that the main computation overhead switches from matrix multiplication to the data transfer between FPGA and HBM. In Figure~\ref{Fig:res_replace}, we show the changes in memorization speed and FPGA HBM data communication with different policies and memory usage. In Figure~\ref{Fig:res_replace}, those UltraRAMs are only used to store vertex hypervectors ($\textbf{H}^\textbf{v}$). Due to the limited on-chip storage, the FPGA will need to replace hypervectors currently stored on-chip with what the incoming ones needed for performing aggregation operations. This process is like a cache miss in the CPU memory hierarchy. Those access misses will bring unnecessary data communication between FPGA and HBM. As is shown in Figure~\ref{Fig:res_replace}, when on-chip UltraRAM usage is small (such as 64 UltraRAM), then data transfer between FPGA and HBM is large. As expected, with the increased usage of on-chip UltraRAM, both the memorization time and FPGA-HBM communication time decrease. This trend is especially noticeable for large-scale datasets like YAGO3-10 (Figure~\ref{Fig:res_replace}.(d)).        

In Figure~\ref{Fig:res_replace}, we present the influence of different replacement policies. We tried three different replacement policies including random replacement (\textbf{Random}), last frequently use (\textbf{LFU}), and last recent use (\textbf{LRU}). On average, LFU achieves the best performance and 8\% better than Random. To further extend our findings, future work could involve designing specific replacement policies targeting KG datasets.  

\subsection{Cross Models and Platforms Comparison}
Figure~\ref{Fig:speedup_compare} presents the speedup and energy efficiency comparison between different models on different hardware platforms. 
Specifically, we show 4 different graph models on 8 different hardware platforms. 
For 8 hardware platforms, we try 2 different CPUs (Intel i9-12900 KF and AMD Ryzen 5955 WX), 3 different GPUs (NVIDIA RTX 3090, RTX 4090, and A100), and 5 different FPGAs (Xilinx Alveo U50, U200, U250, U280, and Kintex7 KC705). On the GPU platforms, we choose to use PyG to optimize models' execution speed and GPU memory usage. 

For FPGA platforms, the configuration of our design on Alveo U50 is the same as Table~\ref{tab:FPGA_resource}. For Alveo U280, with more resources, we increase the HBM channels from 8 PCs to 16 PCs, set the AXI data width as 512, increase the number of Memorization Computing IP ($\textbf{N}_\textbf{c}$) from 16 to 32, training chunk size (\textbf{T}) from 32 to 64, and use 256 UltraRAMs to store vertex hypervectors on-chip. We approximate the HDR's performance on existing state-of-the-art HDC FPGA accelerator (LookHD)~\cite{imani2021revisiting}. We also approximate the performance of 4 different models on Alveo U200 and U250 based on state-of-the-art GNN training FPGA acceleration works (GraphACT~\cite{zeng2020graphact} and HP-GNN~\cite{lin2022hp}).

To ensure a fair comparison, we compare the small FPGA design (Alveo U50) with GraphACT which is based on Alveo U200, and compare the large FPGA design (Alveo U280) with HP-GNN which is based on Alveo U250. 
Compared to GraphACT (Alveo U200), \Design on Alveo U50 shows on average 9$\times$ speedup and 10$\times$ energy efficiency improvement. When comparing with HP-GNN (Alveo U250), \Design on Alveo U280 shows on average 3.5$\times$ speedup and 4.6$\times$ energy efficiency improvement. 
When comparing with RTX 4090, our Alveo U280 accelerator achieves on average \textbf{10.6$\times$} speedup and \textbf{65$\times$} more energy efficiency. 
When conducting cross models and platforms comparison, \Design on Xilinx Alveo U280 (and Xilinx Alveo U50) provides on average 4.2$\times$ (and 8.3$\times$) speedup and 3.4$\times$ (and 9.1$\times$) energy efficiency improvement with similar accuracy as compared to the GCN training platform HP-GNN (and GraphACT).      

\section{Related Works}
\noindent \textbf{Knowledge Graph Completion (KGC)} In contrast to conventional graph learning applications like vertex and graph classification, KGC allows for the derivation of new knowledge and insights from pre-existing data~\cite{chen2020review}. Prior works focused on KG reasoning acceleration have predominantly employed graph embedding-based algorithms~\cite{bordes2013translating,sun2019rotate,yang2014embedding}. These have been accelerated using both CPU~\cite{zhu2019graphvite,arm2020torchkge} and GPU~\cite{zheng2020dgl} platforms. Nevertheless, embedding-based KG reasoning algorithms tend to struggle with achieving high prediction accuracy for more complex KGs. To overcome this limitation and enhance reasoning accuracy, SOTA KG reasoning models attempt to combine embedding models with graph convolution neural networks (GCN)~\cite{vashishth2019composition,shang2019end,schlichtkrull2018modeling}. However, the training overhead associated with GCN-based models is significant on existing hardware platforms. 

\noindent \textbf{GCN Accelerators} In recent years, domain-specific accelerators (DSAs) targeting the graph convolution neural network (GCN) model have been highly successful at top system and architecture conferences. Prior works have primarily focused on accelerating GCN inference~\cite{yan2020hygcn,li2021gcnax,you2022gcod,huang2022accelerating,li2022hyperscale,geng2020awb,zhou2022model0,zhang2022decgnn,sarkar2023flowgnn}, as well as graph neural network (GNN) training~\cite{zeng2020graphact,lin2022hp,chen2021rubik}. However, these earlier efforts focus on traditional tasks such as vertex and graph classification, with little emphasis placed on KGC applications. 

\noindent \textbf{Hyperdimension Computing} In recent years, hyperdimensional computing (HDC) has exhibited significant potential in learning applications 
surpassing conventional machine learning methods like deep neural networks (DNN)~\cite{imani2020dual,zou2022biohd,jiao2022brain,chang2023recent,ge2020classification,imani2021revisiting,HDPG_Yang,ni2022algorithm}. In particular, HDC has demonstrated substantial promise in competing with graph convolution neural networks (GCN) in the graph learning domain~\cite{poduval2022graphd,nunes2022graphhd,kang2022relhd}. Despite this progress, previous HDC-based works have not endeavored to undertake KGC tasks. 

\section{Conclusion}
In this paper, we conduct algorithm and hardware co-design and propose the first HDC-based KGC acceleration platform named \Design. Regarding algorithm, \Design achieves relatively high accuracy when compared to state-of-the-art embedding and GCN models, as demonstrated through experiments using real-life KG datasets. Regarding hardware, the paper describes the design of an FPGA accelerator with multiple optimizations targeting \Design model. 
\Design on Xilinx Alveo U280 provides, on average, a \textbf{4.2$\times$} speedup and \textbf{3.4$\times$} energy efficiency improvement with similar accuracy as compared to the GCN training platform when conducting cross-model and platform comparisons.

\section*{Acknowledgements}
This work was supported in part by the DARPA Young Faculty Award, the National Science Foundation (NSF) under Grants \#2127780, \#2319198, \#2321840, \#2312517, and \#2235472, the Semiconductor Research Corporation (SRC), the Office of Naval Research through the Young Investigator Program Award, and Grants \#N00014-21-1-2225 and N00014-22-1-2067. Additionally, support was provided by the Air Force Office of Scientific Research under Award \#FA9550-22-1-0253, along with generous gifts from Xilinx and Cisco.


\newpage
\bibliographystyle{IEEEtranS}
\bibliography{refs}

\begin{thebibliography}{10}
\providecommand{\url}[1]{#1}
\csname url@samestyle\endcsname
\providecommand{\newblock}{\relax}
\providecommand{\bibinfo}[2]{#2}
\providecommand{\BIBentrySTDinterwordspacing}{\spaceskip=0pt\relax}
\providecommand{\BIBentryALTinterwordstretchfactor}{4}
\providecommand{\BIBentryALTinterwordspacing}{\spaceskip=\fontdimen2\font plus
\BIBentryALTinterwordstretchfactor\fontdimen3\font minus \fontdimen4\font\relax}
\providecommand{\BIBforeignlanguage}[2]{{%
\expandafter\ifx\csname l@#1\endcsname\relax
\typeout{** WARNING: IEEEtranS.bst: No hyphenation pattern has been}%
\typeout{** loaded for the language `#1'. Using the pattern for}%
\typeout{** the default language instead.}%
\else
\language=\csname l@#1\endcsname
\fi
#2}}
\providecommand{\BIBdecl}{\relax}
\BIBdecl

\bibitem{bordes2013translating}
A.~Bordes, N.~Usunier, A.~Garcia-Duran, J.~Weston, and O.~Yakhnenko, ``Translating embeddings for modeling multi-relational data,'' \emph{Advances in neural information processing systems}, vol.~26, 2013.

\bibitem{arm2020torchkge}
A.~Boschin, ``Torchkge: Knowledge graph embedding in python and pytorch,'' in \emph{International Workshop on Knowledge Graph: Mining Knowledge Graph for Deep Insights}, Aug 2020.

\bibitem{chang2023recent}
C.-Y. Chang, Y.-C. Chuang, C.-T. Huang, and A.-Y. Wu, ``Recent progress and development of hyperdimensional computing (hdc) for edge intelligence,'' \emph{IEEE Journal on Emerging and Selected Topics in Circuits and Systems}, 2023.

\bibitem{Hanning_ICCAD2022}
H.~Chen, M.~Issa, Y.~Ni, and M.~Imani, ``Darl: Distributed reconfigurable accelerator for hyperdimensional reinforcement learning,'' in \emph{Proceedings of the 41st IEEE/ACM International Conference on Computer-Aided Design}, 2022, pp. 1--9.

\bibitem{chen2021rubik}
X.~Chen, Y.~Wang, X.~Xie, X.~Hu, A.~Basak, L.~Liang, M.~Yan, L.~Deng, Y.~Ding, Z.~Du, Y.~Chen, and Y.~Xie, ``Rubik: A hierarchical architecture for efficient graph neural network training,'' \emph{IEEE Transactions on Computer-Aided Design of Integrated Circuits and Systems}, vol.~41, no.~4, pp. 936--949, 2021.

\bibitem{chen2020review}
X.~Chen, S.~Jia, and Y.~Xiang, ``A review: Knowledge reasoning over knowledge graph,'' \emph{Expert Systems with Applications}, vol. 141, p. 112948, 2020.

\bibitem{cho2019fa3c}
H.~Cho, P.~Oh, J.~Park, W.~Jung, and J.~Lee, ``Fa3c: Fpga-accelerated deep reinforcement learning,'' in \emph{Proceedings of the Twenty-Fourth International Conference on Architectural Support for Programming Languages and Operating Systems}, 2019, pp. 499--513.

\bibitem{das2017go}
R.~Das, S.~Dhuliawala, M.~Zaheer, L.~Vilnis, I.~Durugkar, A.~Krishnamurthy, A.~Smola, and A.~McCallum, ``Go for a walk and arrive at the answer: Reasoning over paths in knowledge bases using reinforcement learning,'' \emph{arXiv preprint arXiv:1711.05851}, 2017.

\bibitem{dettmers2018convolutional}
T.~Dettmers, P.~Minervini, P.~Stenetorp, and S.~Riedel, ``Convolutional 2d knowledge graph embeddings,'' in \emph{Proceedings of the AAAI conference on artificial intelligence}, vol.~32, no.~1, 2018.

\bibitem{feist2012vivado}
T.~Feist, ``Vivado design suite,'' \emph{White Paper}, vol.~5, p.~30, 2012.

\bibitem{fey2019fast}
M.~Fey and J.~E. Lenssen, ``Fast graph representation learning with pytorch geometric,'' \emph{arXiv preprint arXiv:1903.02428}, 2019.

\bibitem{gao2021quatde}
H.~Gao, K.~Yang, Y.~Yang, R.~Y. Zakari, J.~W. Owusu, and K.~Qin, ``Quatde: Dynamic quaternion embedding for knowledge graph completion,'' \emph{arXiv preprint arXiv:2105.09002}, 2021.

\bibitem{ge2020classification}
L.~Ge and K.~K. Parhi, ``Classification using hyperdimensional computing: A review,'' \emph{IEEE Circuits and Systems Magazine}, vol.~20, no.~2, pp. 30--47, 2020.

\bibitem{geng2020awb}
T.~Geng, A.~Li, R.~Shi, C.~Wu, T.~Wang, Y.~Li, P.~Haghi, A.~Tumeo, S.~Che, S.~Reinhardt, and M.~C. Herbordt, ``Awb-gcn: A graph convolutional network accelerator with runtime workload rebalancing,'' in \emph{2020 53rd Annual IEEE/ACM International Symposium on Microarchitecture (MICRO)}.\hskip 1em plus 0.5em minus 0.4em\relax IEEE, 2020, pp. 922--936.

\bibitem{guo2020survey}
Q.~Guo, F.~Zhuang, C.~Qin, H.~Zhu, X.~Xie, H.~Xiong, and Q.~He, ``A survey on knowledge graph-based recommender systems,'' \emph{IEEE Transactions on Knowledge and Data Engineering}, vol.~34, no.~8, pp. 3549--3568, 2020.

\bibitem{hildebrandt2020reasoning}
M.~Hildebrandt, J.~A.~Q. Serna, Y.~Ma, M.~Ringsquandl, M.~Joblin, and V.~Tresp, ``Reasoning on knowledge graphs with debate dynamics,'' \emph{arXiv preprint arXiv:2001.00461}, 2020.

\bibitem{huang2019knowledge}
X.~Huang, J.~Zhang, D.~Li, and P.~Li, ``Knowledge graph embedding based question answering,'' in \emph{Proceedings of the twelfth ACM international conference on web search and data mining}, 2019, pp. 105--113.

\bibitem{huang2022accelerating}
Y.~Huang, L.~Zheng, P.~Yao, Q.~Wang, X.~Liao, H.~Jin, and J.~Xue, ``Accelerating graph convolutional networks using crossbar-based processing-in-memory architectures,'' in \emph{2022 IEEE International Symposium on High-Performance Computer Architecture (HPCA)}.\hskip 1em plus 0.5em minus 0.4em\relax IEEE, 2022, pp. 1029--1042.

\bibitem{imani2017voicehd}
M.~Imani, D.~Kong, A.~Rahimi, and T.~Rosing, ``Voicehd: Hyperdimensional computing for efficient speech recognition,'' in \emph{2017 IEEE international conference on rebooting computing (ICRC)}.\hskip 1em plus 0.5em minus 0.4em\relax IEEE, 2017, pp. 1--8.

\bibitem{imani2020dual}
M.~Imani, S.~Pampana, S.~Gupta, M.~Zhou, Y.~Kim, and T.~Rosing, ``Dual: Acceleration of clustering algorithms using digital-based processing in-memory,'' in \emph{2020 53rd Annual IEEE/ACM International Symposium on Microarchitecture (MICRO)}.\hskip 1em plus 0.5em minus 0.4em\relax IEEE, 2020, pp. 356--371.

\bibitem{imani2019sparsehd}
M.~Imani, S.~Salamat, B.~Khaleghi, M.~Samragh, F.~Koushanfar, and T.~Rosing, ``Sparsehd: Algorithm-hardware co-optimization for efficient high-dimensional computing,'' in \emph{2019 IEEE 27th Annual International Symposium on Field-Programmable Custom Computing Machines (FCCM)}.\hskip 1em plus 0.5em minus 0.4em\relax IEEE, 2019, pp. 190--198.

\bibitem{imani2021revisiting}
M.~Imani, Z.~Zou, S.~Bosch, S.~A. Rao, S.~Salamat, V.~Kumar, Y.~Kim, and T.~Rosing, ``Revisiting hyperdimensional learning for fpga and low-power architectures,'' in \emph{2021 IEEE International Symposium on High-Performance Computer Architecture (HPCA)}.\hskip 1em plus 0.5em minus 0.4em\relax IEEE, 2021, pp. 221--234.

\bibitem{jiao2022brain}
X.~Jiao, A.~Rahimi, C.~Ferm{\"u}ller, and J.~Y. Aloimonos, ``Brain-inspired hyperdimensional computing: Algorithms, models, and architectures,'' \emph{Frontiers in Neuroscience}, vol.~16, 2022.

\bibitem{kang2022relhd}
J.~Kang, M.~Zhou, A.~Bhansali, W.~Xu, A.~Thomas, and T.~Rosing, ``Relhd: A graph-based learning on fefet with hyperdimensional computing,'' in \emph{2022 IEEE 40th International Conference on Computer Design (ICCD)}.\hskip 1em plus 0.5em minus 0.4em\relax IEEE, 2022, pp. 553--560.

\bibitem{karunaratne2020memory}
G.~Karunaratne, M.~Le~Gallo, G.~Cherubini, L.~Benini, A.~Rahimi, and A.~Sebastian, ``In-memory hyperdimensional computing,'' \emph{Nature Electronics}, vol.~3, no.~6, pp. 327--337, 2020.

\bibitem{kathail2020xilinx}
V.~Kathail, ``Xilinx vitis unified software platform,'' in \emph{Proceedings of the 2020 ACM/SIGDA International Symposium on Field-Programmable Gate Arrays}, 2020, pp. 173--174.

\bibitem{kiningham2022grip}
K.~Kiningham, P.~Levis, and C.~R{\'e}, ``Grip: A graph neural network accelerator architecture,'' \emph{IEEE Transactions on Computers}, vol.~72, no.~4, pp. 914--925, 2022.

\bibitem{kleyko2023survey}
D.~Kleyko, D.~Rachkovskij, E.~Osipov, and A.~Rahimi, ``A survey on hyperdimensional computing aka vector symbolic architectures, part ii: Applications, cognitive models, and challenges,'' \emph{ACM Computing Surveys}, vol.~55, no.~9, pp. 1--52, 2023.

\bibitem{lacey2016deep}
G.~Lacey, G.~W. Taylor, and S.~Areibi, ``Deep learning on fpgas: Past, present, and future,'' \emph{arXiv preprint arXiv:1602.04283}, 2016.

\bibitem{latapie2021metamodel}
H.~Latapie, O.~Kilic, G.~Liu, R.~Kompella, A.~Lawrence, Y.~Sun, J.~Srinivasa, Y.~Yan, P.~Wang, and K.~R. Th{\'o}risson, ``A metamodel and framework for artificial general intelligence from theory to practice,'' \emph{Journal of Artificial Intelligence and Consciousness}, vol.~8, no.~02, pp. 205--227, 2021.

\bibitem{lee1999existence}
D.~Lee, J.~Choi, J.-H. Kim, S.~H. Noh, S.~L. Min, Y.~Cho, and C.~S. Kim, ``On the existence of a spectrum of policies that subsumes the least recently used (lru) and least frequently used (lfu) policies,'' in \emph{Proceedings of the 1999 ACM SIGMETRICS international conference on Measurement and modeling of computer systems}, 1999, pp. 134--143.

\bibitem{li2021gcnax}
J.~Li, A.~Louri, A.~Karanth, and R.~Bunescu, ``Gcnax: A flexible and energy-efficient accelerator for graph convolutional neural networks,'' in \emph{2021 IEEE International Symposium on High-Performance Computer Architecture (HPCA)}.\hskip 1em plus 0.5em minus 0.4em\relax IEEE, 2021, pp. 775--788.

\bibitem{li2022hyperscale}
S.~Li, D.~Niu, Y.~Wang, W.~Han, Z.~Zhang, T.~Guan, Y.~Guan, H.~Liu, L.~Huang, Z.~Du, F.~Xue, Y.~Fang, H.~Zheng, and X.~Yuan, ``Hyperscale fpga-as-a-service architecture for large-scale distributed graph neural network,'' in \emph{Proceedings of the 49th Annual International Symposium on Computer Architecture}, 2022, pp. 946--961.

\bibitem{lin2022hp}
Y.-C. Lin, B.~Zhang, and V.~Prasanna, ``Hp-gnn: generating high throughput gnn training implementation on cpu-fpga heterogeneous platform,'' in \emph{Proceedings of the 2022 ACM/SIGDA International Symposium on Field-Programmable Gate Arrays}, 2022, pp. 123--133.

\bibitem{mahdisoltani2014yago3}
F.~Mahdisoltani, J.~Biega, and F.~Suchanek, ``Yago3: A knowledge base from multilingual wikipedias,'' in \emph{7th biennial conference on innovative data systems research}.\hskip 1em plus 0.5em minus 0.4em\relax CIDR Conference, 2014.

\bibitem{miller1995wordnet}
G.~A. Miller, ``Wordnet: a lexical database for english,'' \emph{Communications of the ACM}, vol.~38, no.~11, pp. 39--41, 1995.

\bibitem{montagna2018pulp}
F.~Montagna, A.~Rahimi, S.~Benatti, D.~Rossi, and L.~Benini, ``Pulp-hd: Accelerating brain-inspired high-dimensional computing on a parallel ultra-low power platform,'' in \emph{Proceedings of the 55th Annual Design Automation Conference}, 2018, pp. 1--6.

\bibitem{ni2023brain}
Y.~Ni, H.~Chen, P.~Poduval, Z.~Zou, P.~Mercati, and M.~Imani, ``Brain-inspired trustworthy hyperdimensional computing with efficient uncertainty quantification,'' in \emph{2023 IEEE/ACM International Conference on Computer Aided Design (ICCAD)}.\hskip 1em plus 0.5em minus 0.4em\relax IEEE, 2023, pp. 01--09.

\bibitem{HDPG_Yang}
\BIBentryALTinterwordspacing
Y.~Ni, M.~Issa, D.~Abraham, M.~Imani, X.~Yin, and M.~Imani, ``Hdpg: Hyperdimensional policy-based reinforcement learning for continuous control,'' in \emph{Proceedings of the 59th ACM/IEEE Design Automation Conference}, ser. DAC '22, 2022, p. 1141–1146. [Online]. Available: \url{https://doi.org/10.1145/3489517.3530668}
\BIBentrySTDinterwordspacing

\bibitem{ni2022algorithm}
Y.~Ni, Y.~Kim, T.~Rosing, and M.~Imani, ``Algorithm-hardware co-design for efficient brain-inspired hyperdimensional learning on edge,'' in \emph{2022 Design, Automation \& Test in Europe Conference \& Exhibition (DATE)}.\hskip 1em plus 0.5em minus 0.4em\relax IEEE, 2022, pp. 292--297.

\bibitem{ni2022neurally}
Y.~Ni, N.~Lesica, F.-G. Zeng, and M.~Imani, ``Neurally-inspired hyperdimensional classification for efficient and robust biosignal processing,'' in \emph{Proceedings of the 41st IEEE/ACM International Conference on Computer-Aided Design}, 2022, pp. 1--9.

\bibitem{nunes2022graphhd}
I.~Nunes, M.~Heddes, T.~Givargis, A.~Nicolau, and A.~Veidenbaum, ``Graphhd: Efficient graph classification using hyperdimensional computing,'' in \emph{2022 Design, Automation \& Test in Europe Conference \& Exhibition (DATE)}.\hskip 1em plus 0.5em minus 0.4em\relax IEEE, 2022, pp. 1485--1490.

\bibitem{paszke2019pytorch}
A.~Paszke, S.~Gross, F.~Massa, A.~Lerer, J.~Bradbury, G.~Chanan, T.~Killeen, Z.~Lin, N.~Gimelshein, L.~Antiga, A.~Desmaison, A.~Köpf, E.~Yang, Z.~DeVito, M.~Raison, A.~Tejani, S.~Chilamkurthy, B.~Steiner, L.~Fang, J.~Bai, and S.~Chintala, ``Pytorch: An imperative style, high-performance deep learning library,'' \emph{Advances in neural information processing systems}, vol.~32, 2019.

\bibitem{poduval2022graphd}
P.~Poduval, H.~Alimohamadi, A.~Zakeri, F.~Imani, M.~H. Najafi, T.~Givargis, and M.~Imani, ``Graphd: Graph-based hyperdimensional memorization for brain-like cognitive learning,'' \emph{Frontiers in Neuroscience}, vol.~16, p.~5, 2022.

\bibitem{F5-HD}
S.~Salamat, M.~Imani, B.~Khaleghi, and T.~Rosing, ``F5-hd: Fast flexible fpga-based framework for refreshing hyperdimensional computing,'' in \emph{Proceedings of the 2019 ACM/SIGDA International Symposium on Field-Programmable Gate Arrays}, 2019, pp. 53--62.

\bibitem{salamat2020accelerating}
S.~Salamat, M.~Imani, and T.~Rosing, ``Accelerating hyperdimensional computing on fpgas by exploiting computational reuse,'' \emph{IEEE Transactions on Computers}, vol.~69, no.~8, pp. 1159--1171, 2020.

\bibitem{sarkar2023flowgnn}
R.~Sarkar, S.~Abi-Karam, Y.~He, L.~Sathidevi, and C.~Hao, ``Flowgnn: A dataflow architecture for real-time workload-agnostic graph neural network inference,'' in \emph{2023 IEEE International Symposium on High-Performance Computer Architecture (HPCA)}.\hskip 1em plus 0.5em minus 0.4em\relax IEEE, 2023, pp. 1099--1112.

\bibitem{10071015}
R.~Sarkar, S.~Abi-Karam, Y.~He, L.~Sathidevi, and C.~Hao, ``Flowgnn: A dataflow architecture for real-time workload-agnostic graph neural network inference,'' in \emph{2023 IEEE International Symposium on High-Performance Computer Architecture (HPCA)}, 2023, pp. 1099--1112.

\bibitem{schlichtkrull2018modeling}
M.~Schlichtkrull, T.~N. Kipf, P.~Bloem, R.~Van Den~Berg, I.~Titov, and M.~Welling, ``Modeling relational data with graph convolutional networks,'' in \emph{The Semantic Web: 15th International Conference, ESWC 2018, Heraklion, Crete, Greece, June 3--7, 2018, Proceedings 15}.\hskip 1em plus 0.5em minus 0.4em\relax Springer, 2018, pp. 593--607.

\bibitem{shang2019end}
C.~Shang, Y.~Tang, J.~Huang, J.~Bi, X.~He, and B.~Zhou, ``End-to-end structure-aware convolutional networks for knowledge base completion,'' in \emph{Proceedings of the AAAI conference on artificial intelligence}, vol.~33, no.~01, 2019, pp. 3060--3067.

\bibitem{song2022sextans}
L.~Song, Y.~Chi, A.~Sohrabizadeh, Y.-k. Choi, J.~Lau, and J.~Cong, ``Sextans: A streaming accelerator for general-purpose sparse-matrix dense-matrix multiplication,'' in \emph{Proceedings of the 2022 ACM/SIGDA International Symposium on Field-Programmable Gate Arrays}, 2022, pp. 65--77.

\bibitem{stoica2020contextual}
G.~Stoica, O.~Stretcu, E.~A. Platanios, T.~Mitchell, and B.~P{\'o}czos, ``Contextual parameter generation for knowledge graph link prediction,'' in \emph{Proceedings of the AAAI Conference on Artificial Intelligence}, vol.~34, no.~03, 2020, pp. 3000--3008.

\bibitem{sun2019rotate}
Z.~Sun, Z.-H. Deng, J.-Y. Nie, and J.~Tang, ``Rotate: Knowledge graph embedding by relational rotation in complex space,'' \emph{arXiv preprint arXiv:1902.10197}, 2019.

\bibitem{toutanova2015observed}
K.~Toutanova and D.~Chen, ``Observed versus latent features for knowledge base and text inference,'' in \emph{Proceedings of the 3rd workshop on continuous vector space models and their compositionality}, 2015, pp. 57--66.

\bibitem{vashishth2019composition}
S.~Vashishth, S.~Sanyal, V.~Nitin, and P.~Talukdar, ``Composition-based multi-relational graph convolutional networks,'' \emph{arXiv preprint arXiv:1911.03082}, 2019.

\bibitem{wang2019knowledge}
H.~Wang, M.~Zhao, X.~Xie, W.~Li, and M.~Guo, ``Knowledge graph convolutional networks for recommender systems,'' in \emph{The world wide web conference}, 2019, pp. 3307--3313.

\bibitem{wang2020adrl}
Q.~Wang, Y.~Hao, and J.~Cao, ``Adrl: An attention-based deep reinforcement learning framework for knowledge graph reasoning,'' \emph{Knowledge-Based Systems}, vol. 197, p. 105910, 2020.

\bibitem{wang2021gnnadvisor}
Y.~Wang, B.~Feng, G.~Li, S.~Li, L.~Deng, Y.~Xie, and Y.~Ding, ``Gnnadvisor: An adaptive and efficient runtime system for gnn acceleration on gpus,'' in \emph{15th USENIX symposium on operating systems design and implementation (OSDI 21)}, 2021.

\bibitem{openmem}
F.~Wen, M.~Qin, P.~Gratz, and N.~Reddy, ``Openmem: Hardware/software cooperative management for mobile memory system,'' in \emph{2021 58th ACM/IEEE Design Automation Conference (DAC)}, 2021, pp. 109--114.

\bibitem{softhint}
F.~Wen, M.~Qin, P.~Gratz, and N.~Reddy, ``Software hint-driven data management for hybrid memory in mobile systems,'' \emph{ACM Trans. Embed. Comput. Syst.}, vol.~21, no.~1, jan 2022.

\bibitem{hmmu}
F.~Wen, M.~Qin, P.~V. Gratz, and A.~L.~N. Reddy, ``Hardware memory management for future mobile hybrid memory systems,'' \emph{IEEE Transactions on Computer-Aided Design of Integrated Circuits and Systems}, vol.~39, no.~11, pp. 3627--3637, 2020.

\bibitem{wu2021findings}
C.~Wu, F.~Wu, and Y.~Huang, ``Findings of the association for computational linguistics: Acl-ijcnlp 2021,'' in \emph{Association for Computational Linguistics}, 2021, pp. 4408--4413.

\bibitem{xiong2017deeppath}
W.~Xiong, T.~Hoang, and W.~Y. Wang, ``Deeppath: A reinforcement learning method for knowledge graph reasoning,'' \emph{arXiv preprint arXiv:1707.06690}, 2017.

\bibitem{yan2020characterizing}
M.~Yan, Z.~Chen, L.~Deng, X.~Ye, Z.~Zhang, D.~Fan, and Y.~Xie, ``Characterizing and understanding gcns on gpu,'' \emph{IEEE Computer Architecture Letters}, vol.~19, no.~1, pp. 22--25, 2020.

\bibitem{yan2020hygcn}
M.~Yan, L.~Deng, X.~Hu, L.~Liang, Y.~Feng, X.~Ye, Z.~Zhang, D.~Fan, and Y.~Xie, ``Hygcn: A gcn accelerator with hybrid architecture,'' in \emph{2020 IEEE International Symposium on High Performance Computer Architecture (HPCA)}.\hskip 1em plus 0.5em minus 0.4em\relax IEEE, 2020, pp. 15--29.

\bibitem{yang2015embedding}
B.~Yang, S.~W.-t. Yih, X.~He, J.~Gao, and L.~Deng, ``Embedding entities and relations for learning and inference in knowledge bases,'' in \emph{Proceedings of the International Conference on Learning Representations (ICLR) 2015}, 2015.

\bibitem{yang2014embedding}
B.~Yang, W.-t. Yih, X.~He, J.~Gao, and L.~Deng, ``Embedding entities and relations for learning and inference in knowledge bases,'' \emph{arXiv preprint arXiv:1412.6575}, 2014.

\bibitem{yang2022gnnlab}
J.~Yang, D.~Tang, X.~Song, L.~Wang, Q.~Yin, R.~Chen, W.~Yu, and J.~Zhou, ``Gnnlab: a factored system for sample-based gnn training over gpus,'' in \emph{Proceedings of the Seventeenth European Conference on Computer Systems}, 2022, pp. 417--434.

\bibitem{you2022gcod}
H.~You, T.~Geng, Y.~Zhang, A.~Li, and Y.~Lin, ``Gcod: Graph convolutional network acceleration via dedicated algorithm and accelerator co-design,'' in \emph{2022 IEEE International Symposium on High-Performance Computer Architecture (HPCA)}.\hskip 1em plus 0.5em minus 0.4em\relax IEEE, 2022, pp. 460--474.

\bibitem{zeng2020graphact}
H.~Zeng and V.~Prasanna, ``Graphact: Accelerating gcn training on cpu-fpga heterogeneous platforms,'' in \emph{Proceedings of the 2020 ACM/SIGDA International Symposium on Field-Programmable Gate Arrays}, 2020, pp. 255--265.

\bibitem{zeng2022toward}
X.~Zeng, X.~Tu, Y.~Liu, X.~Fu, and Y.~Su, ``Toward better drug discovery with knowledge graph,'' \emph{Current opinion in structural biology}, vol.~72, pp. 114--126, 2022.

\bibitem{zhang2022decgnn}
B.~Zhang, H.~Zeng, and V.~K. Prasanna, ``Decgnn: A framework for mapping decoupled gnn models onto cpu-fpga heterogeneous platform,'' in \emph{Proceedings of the 2022 ACM/SIGDA International Symposium on Field-Programmable Gate Arrays}, 2022, pp. 154--154.

\bibitem{zhang2019qpytorch}
T.~Zhang, Z.~Lin, G.~Yang, and C.~De~Sa, ``Qpytorch: A low-precision arithmetic simulation framework,'' in \emph{2019 Fifth Workshop on Energy Efficient Machine Learning and Cognitive Computing-NeurIPS Edition (EMC2-NIPS)}.\hskip 1em plus 0.5em minus 0.4em\relax IEEE, 2019, pp. 10--13.

\bibitem{zhang2018variational}
Y.~Zhang, H.~Dai, Z.~Kozareva, A.~Smola, and L.~Song, ``Variational reasoning for question answering with knowledge graph,'' in \emph{Proceedings of the AAAI conference on artificial intelligence}, vol.~32, no.~1, 2018.

\bibitem{zheng2020dgl}
D.~Zheng, X.~Song, C.~Ma, Z.~Tan, Z.~Ye, J.~Dong, H.~Xiong, Z.~Zhang, and G.~Karypis, ``Dgl-ke: Training knowledge graph embeddings at scale,'' in \emph{Proceedings of the 43rd International ACM SIGIR Conference on Research and Development in Information Retrieval}, 2020, pp. 739--748.

\bibitem{zhou2022model0}
H.~Zhou, B.~Zhang, R.~Kannan, V.~Prasanna, and C.~Busart, ``Model-architecture co-design for high performance temporal gnn inference on fpga,'' \emph{arXiv preprint arXiv:2203.05095}, 2022.

\bibitem{zhu2019graphvite}
Z.~Zhu, S.~Xu, M.~Qu, and J.~Tang, ``Graphvite: A high-performance cpu-gpu hybrid system for node embedding,'' in \emph{The World Wide Web Conference}.\hskip 1em plus 0.5em minus 0.4em\relax ACM, 2019, pp. 2494--2504.

\bibitem{zou2022biohd}
Z.~Zou, H.~Chen, P.~Poduval, Y.~Kim, M.~Imani, E.~Sadredini, R.~Cammarota, and M.~Imani, ``Biohd: an efficient genome sequence search platform using hyperdimensional memorization,'' in \emph{Proceedings of the 49th Annual International Symposium on Computer Architecture}, 2022, pp. 656--669.

\bibitem{zou2021scalable}
Z.~Zou, Y.~Kim, F.~Imani, H.~Alimohamadi, R.~Cammarota, and M.~Imani, ``Scalable edge-based hyperdimensional learning system with brain-like neural adaptation,'' in \emph{Proceedings of the International Conference for High Performance Computing, Networking, Storage and Analysis}, 2021, pp. 1--15.

\end{thebibliography}

\end{document}